\pdfoutput=1
\documentclass[cernpreprint, atlasdraft=false, texlive=2023, UKenglish, orcidlogo, texmf, orcidlogo]{atlasdoc}
 
\usepackage[block=space]{atlaspackage}
\usepackage{atlasbiblatex}
 
\usepackage{atlasphysics}
 
\graphicspath{{logos/}{figures/}}
 
\usepackage{ANA-SUSY-2018-11-PAPER-defs}

\AtlasTitle{Search for new phenomena in final states with photons, jets and missing transverse momentum in $pp$ collisions at $\sqrt{s} = 13$~\ensuremath{\text{Te\kern -0.1em V}} with the ATLAS detector}

\AtlasAbstract{
 
A search for new phenomena has been performed in final states with at least one isolated high-momentum photon, jets and missing transverse momentum in proton--proton collisions at a centre-of-mass energy of \ensuremath{\sqrt{s} = 13 \; \text{Te\kern -0.1em V}}. The data, collected by the ATLAS experiment at the CERN LHC, correspond to an integrated luminosity of \ensuremath{139 \; \mbox{fb\(^{-1}\)}}. The experimental results are interpreted in a supersymmetric model in which pair-produced gluinos decay into neutralinos, which in turn decay into a gravitino, at least one photon, and jets.
No significant deviations from the predictions of the Standard Model are observed. Upper limits are set on the visible cross section due to physics beyond the Standard Model, and lower limits are set on the masses of the gluinos and neutralinos, all at 95\% confidence level. Visible cross sections greater than 0.022~fb are excluded and pair-produced gluinos with masses up to 2200~\gev\ are excluded for most of the NLSP masses investigated.
}

\PreprintIdNumber{CERN-EP-2022-012}

\AtlasJournalRef{\JHEP 07 (2023) 021}
\AtlasDOI{10.1007/JHEP07(2023)021}

\hypersetup{pdftitle={ATLAS document},pdfauthor={The ATLAS Collaboration}}
 
\begin{document}
 
\maketitle

\section{Introduction}
\label{sec:intro}

 
In this paper, proton--proton collisions at the LHC are used to search for new phenomena in experimental signatures with photons, jets and a large amount of missing transverse momentum in the final state.
These signatures are motivated by a gauge-mediated supersymmetry breaking (GMSB) model~\cite{Dine:1981gu,AlvarezGaume:1981wy,Nappi:1982hm}, and its more generalized form, general gauge mediation (GGM) \cite{Cheung:2007es,Meade:2008wd}, where supersymmetry (SUSY) breaking takes place in a hidden sector
and communicates with the visible sector through Standard Model (SM) gauge boson interactions. The results are interpreted in the context of a set of simpified GGM models
that include the production of gluinos (\gluino) -- supersymmetric partners of strongly coupled SM particles - and various assumptions for the couplings of the new particles to the SM bosons.
 
In this scenario, the lightest supersymmetric particle (LSP) is the ultralight gravitino (\gravino), which passes through the detector undetected and induces a non-zero missing transverse momentum (\MET) in the events in which it is produced. In certain circumstances it is a viable dark-matter candidate \cite{Ellis:1983ew,Viel_2005,Kamada2014}.
The models considered in this analysis conserve R-parity~\cite{Farrar:1978xj}, so supersymmetric particles are produced in pairs and each decay must contain an odd number of supersymmetric decay products.
Each decay therefore results in a decay chain leading to the LSP, usually proceeding via the emission of jets through the next-to-lightest supersymmetric particle (NLSP), which is often the lightest neutralino (\ninoone).
 
In the simplified GGM models considered, the decoupled mass scales for supersymmetric partners of the SM particles allow the NLSP neutralino to have large higgsino or bino components. This kind of NLSP decays into a gravitino and a photon or a gravitino and either a $Z$ boson or a Higgs boson ($h$, assumed to have a mass of 125~\gev\ with SM-like couplings). Thus, GGM models with a neutralino NLSP predict final states with two of these bosons (photon, $Z$, or $h$) and two LSPs, and hence large \MET. The final state targeted by this search corresponds to a signature including many jets from the decay chain and \met\ from the undetected particles, in combination with a high transverse momentum (\pt) photon appearing in the decay because of the bino component of the NLSP allowed by GGM.

Examples of production modes in proton--proton collisions for the topologies (from now on called $\gamma/Z$ and $\gamma/h$) targeted in this paper are shown in Figure~\ref{fig:GGM_diagrams}, where all decays modes for the $Z$ and $h$ bosons are considered.

An event selection strategy is designed to maximize the sensitivity for final states with photons, jets and \MET, with the aim of avoiding the imposition of any model-tailored selection requirements.
 
Previous studies of similar signatures with smaller datasets (\prevlumi), collected between 2015 and 2016 at \COMtev, were presented in Refs.~\cite{SUSY-2014-01,SUSY-2016-27}. 
This analysis extends the reach for gluino masses by several hundred \GeV\ by using new criteria for reconstructed objects, new optimized signal regions and background estimation strategies, and the much larger full Run-2 dataset, with an integrated luminosity of 139~\ifb\ recorded during 2015--2018.
The results complement searches \cite{CMS-SUS-17-011,CMS-SUS-18-002,CMS-SUS-18-005} performed by the CMS Collaboration in $pp$ collisions at \COMtev.

\begin{figure}[ht!]
\centering
\subfloat[$\gamma/Z$ model]{\includegraphics[width=0.4\textwidth]{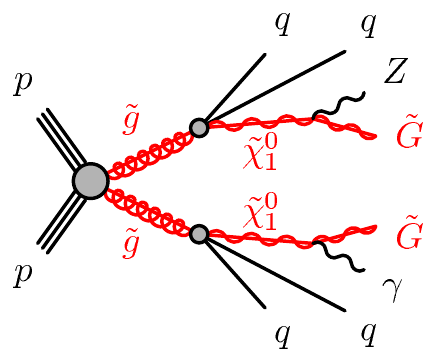}}
\subfloat[$\gamma/h$ model]{\includegraphics[width=0.4\textwidth]{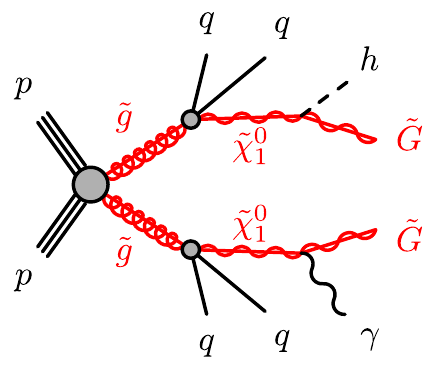}}
\caption{Examples of production diagrams for gluinos and their subsequent decay to a final state with photons, jets and missing transverse momentum (from the gravitinos) for the (a) $\gamma/Z$ and (b) $\gamma/h$ models.}
\label{fig:GGM_diagrams}
\end{figure}


\section{ATLAS detector}
\label{sec:detector}

\newcommand{\AtlasCoordFootnote}{
ATLAS uses a right-handed coordinate system with its origin at the nominal interaction point (IP)
in the centre of the detector and the \(z\)-axis along the beam pipe.
The \(x\)-axis points from the IP to the centre of the LHC ring,
and the \(y\)-axis points upwards.
Cylindrical coordinates \((r,\phi)\) are used in the transverse plane,
\(\phi\) being the azimuthal angle around the \(z\)-axis.
The pseudorapidity is defined in terms of the polar angle \(\theta\) as \(\eta = -\ln \tan(\theta/2)\).
Angular distance is measured in units of \(\Delta R \equiv \sqrt{(\Delta\eta)^{2} + (\Delta\phi)^{2}}\).}
 
The ATLAS detector~\cite{PERF-2007-01} at the LHC covers nearly the entire solid angle around the collision point.\footnote{\AtlasCoordFootnote}
It consists of an inner tracking detector surrounded by a thin superconducting solenoid, electromagnetic and hadron calorimeters,
and a muon spectrometer incorporating three large superconducting air-core toroidal magnets.
 
The inner-detector system (ID) is immersed in a \SI{2}{\tesla} axial magnetic field
and provides charged-particle tracking in the range \(|\eta|< 2.5\).
The high-granularity silicon pixel detector covers the vertex region and typically provides four measurements per track,
the first hit normally being in the insertable B-layer~\cite{ATLAS-TDR-19,PIX-2018-001} installed before Run~2.
It is followed by the silicon microstrip tracker, which usually provides eight measurements per track.
These silicon detectors are complemented by the transition radiation tracker (TRT),
which enables radially extended track reconstruction up to \(|\eta|= 2.0\).
The TRT also provides electron identification information
based on the fraction of hits (typically 30 in total) above a higher energy-deposit threshold corresponding to transition radiation.
 
The calorimeter system covers the pseudorapidity range \(|\eta|< 4.9\).
Within the region \(|\eta|< 3.2\), electromagnetic calorimetry is provided by barrel and
endcap high-granularity lead/liquid-argon (LAr) calorimeters,
with an additional thin LAr presampler covering \(|\eta|< 1.8\)
to correct for energy loss in material upstream of the calorimeters.
Hadron calorimetry is provided by the steel/scintillator-tile calorimeter,
segmented into three barrel structures within \(|\eta|< 1.7\), and two copper/LAr hadron endcap calorimeters.
The solid angle coverage is completed with forward copper/LAr and tungsten/LAr calorimeter modules
optimized for electromagnetic and hadronic energy measurements respectively.
 
The muon spectrometer (MS) comprises separate trigger and
high-precision tracking chambers measuring the deflection of muons in a magnetic field generated by the superconducting air-core toroidal magnets.
The field integral of the toroids ranges between \num{2.0} and \SI{6.0}{\tesla\metre}
across most of the detector.
A set of precision chambers covers the region \(|\eta|< 2.7\) with three layers of monitored drift tubes,
complemented by cathode-strip chambers in the forward region, where the background is highest.
The muon trigger system covers the range \(|\eta|< 2.4\) with resistive-plate chambers in the barrel, and thin-gap chambers in the endcap regions.
 
Interesting events are selected by the first-level trigger system implemented in custom hardware,
followed by selections made by algorithms implemented in software in the high-level trigger~\cite{TRIG-2016-01}.
The first-level trigger accepts events from the \SI{40}{\MHz} bunch crossings at a rate below \SI{100}{\kHz},
which the high-level trigger further reduces in order to record events to disk at about \SI{1}{\kHz}.
 
An extensive software suite~\cite{ATL-SOFT-PUB-2021-001} is used in the reconstruction and analysis of real
and simulated data, in detector operations, and in the trigger and data acquisition systems of the experiment.


\section{Samples of simulated processes}
\label{sec:mc}

Samples of the targeted SUSY signals and SM backgrounds were simulated at $\sqrt{s} = 13~\tev$ using dedicated Monte Carlo (MC) generators.
For the interpretation of the results, a grid of signal samples was simulated with a set of benchmark parameter values covering the region in which the signal can be observed. In these particular regions of the GGM model space, the lightest neutralino is a mixture of bino and higgsino fields; the neutral wino field has a much larger mass than the bino/higgsino, so the corresponding wino content of the lightest neutralino is negligible. The gluino is the only relevant coloured particle, since all squark (supersymmetric partners of the SM quarks) soft masses are decoupled at a value of $5~\tev$. The full model parameters include the U(1), SU(2) and SU(3) gauge partner mass parameters ($M_1$, $M_2$ and $M_3$, respectively), the higgsino mass parameter $\mu$, the gravitino mass $m_{\gravino}$, and the ratio $\tan \beta$ of the two SUSY Higgs-doublet vacuum expectation values. Due to the Weinberg mixing angle in the SM, the bino component of the lightest neutralino couples to both the photon and the $Z$ boson in the case of positive values of $\mu$, and to the photon and the Higgs boson in the case of negative $\mu$ values \cite{Martin:1997ns,Shih2012}.
For all GGM models considered, the phenomenology relevant to this search is only
weakly dependent on the value of $\tan \beta$, chosen to be 1.5. The mass parameter $M_2$ is decoupled at a value of $3~\tev$ and $M_3$ matches the gluino mass since no radiative corrections are taken into account.  The lifetime $\tau$ of the NLSP is set so that $c\tau$ is less than 0.1~mm, ensuring that the neutralino decays promptly, where for most of the mass space it is achieved with a minimal gravitino mass of $m_{\gravino} = 10^{-9}~\gev$.
All trilinear coupling terms are set to zero and the slepton masses are set to $5~\tev$. The Higgs sector is in the decoupling regime at $2~\tev$ (except for the lightest neutral Higgs boson). Setting $M_{1} \sim |\mu| \sim m_{\ninoone}$, the branching ratios of the lightest neutralino to $\gamma$\,+\,$\gravino$ and $Z/h$\,+\,$\gravino$ are approximately constant at 50$\%$, maximizing the production of the final states of interest for this search.

The generated signal samples cover the \gluino--\ninoone mass plane in the range 1400--2600~\gev\ for $m_{\gluino}$ and 150--2600~\gev\ for $m_{\ninoone}$, for each of the targeted models.\footnote{The $m_{\gluino}$ range for the $\gamma/h$ model starts at 1200~\gev\ to match the mass range used in the previous publication \cite{SUSY-2014-01}.} 
The full mass spectrum, the gluino and neutralino branching ratios, and their decay widths were calculated using SUSPECT~v2.43 \cite{Djouadi2007426}, SDECAY~v1.5 \cite{MUHLLEITNER200546} and HDECAY~v3.4 \cite{Djouadi:1997yw}, run as part of the SUSYHIT package v1.5a \cite{Djouadi:2006bz}. All signal samples were then generated at leading order (LO) with up to two additional partons
with \MADGRAPH[5] interfaced to \PYTHIA[8]. 
Signal cross sections are calculated at next-to-leading order (NLO) in the strong coupling constant, adding the resummation of soft gluon emission at next-to-leading-logarithm accuracy~\cite{Beenakker:1996ch,Kulesza:2008jb,Kulesza:2009kq,Beenakker:2009ha,Beenakker:2011fu}.
The nominal cross section and its uncertainty are taken from an envelope of
cross-section predictions using different parton distribution function (PDF) sets and factorization and
renormalization scales, as described in Ref.~\cite{Borschensky:2014cia}.

Most of the backgrounds affecting this search were estimated using control samples selected from data,
defined such that one of the background processes becomes dominant, but otherwise
kinematically similar to the signal region (SR). The extrapolations from these control regions (CRs) to the SRs are based on samples of simulated events.
Samples of \ttbargam\ events were generated with MG5\_aMC{@}NLO
\cite{Alwall:2014hca} at NLO (with $m_{\text{top}}=172.5~\gev$), interfaced to the \PYTHIA[8] parton
shower model \cite{Sjostrand:2014zea}. The \NNPDF[3.0nnlo]~\cite{Ball:2014uwa} set of PDFs was used, with parameter values set to the A14 tune \cite{ATL-PHYS-PUB-2014-021}.
The rest of the SM backgrounds were simulated with the \SHERPA[2.2]~\cite{Bothmann:2019yzt} generator. Matrix elements at LO or NLO accuracy in QCD were calculated with the \comix~\cite{Gleisberg:2008fv} and \openloops\ \cite{Buccioni:2019sur,Cascioli:2011va,Denner:2016kdg} libraries. They were matched and merged with the \sherpa\ parton shower based on Catani--Seymour dipole factorization~\cite{Gleisberg:2008fv,Schumann:2007mg} using the MEPS@NLO
prescription~\cite{Hoeche:2011fd,Hoeche:2012yf,Catani:2001cc,Hoeche:2009rj}. These background samples were generated using the \NNPDF[3.0nnlo] set, along with
the dedicated set of tuned parton-shower parameters developed by the \sherpa\ authors.
For the production of $V\gamma$ ($W\gamma$ and $Z\gamma$) and prompt single-photon ({\gjet}) final states, \SHERPA[2.2.2] was used. Matrix elements at NLO for up to one (two) additional parton emission(s) and at LO accuracy for up to three (four) additional parton emissions were calculated for $V\gamma$ ({\gjet}) samples. Diphoton and $V\gamma\gamma$ ($W\gamma\gamma$ and $Z\gamma\gamma$) events were simulated with \SHERPA[2.2.4]. In this set-up, matrix elements at LO
accuracy in QCD for up to three additional parton emissions were
matched and merged with the \sherpa\ parton shower.

In order to avoid any double counting of events when considering both the $V\gamma$ and $V\gamma\gamma$ samples, duplicated processes were removed, focusing on $V\gamma$ events with a photon from QED final-state radiation.

The SM background samples were passed through a full ATLAS detector simulation \cite{SOFT-2010-01} based on \textsc{Geant4} \cite{Geant4}. The SUSY signal samples were passed through \textsc{ATLFAST-II} \cite{ATL-PHYS-PUB-2010-013}, a fast simulation of the ATLAS detector response based on a parametrization of the performance of the ATLAS EM and hadronic calorimeters \cite{ATL-SOFT-PUB-2014-001} and on \textsc{Geant4} elsewhere.
MC events are reconstructed with the same algorithms used for data. An event-by-event reweighting is applied to all MC samples to realistically model the LHC conditions during data collection. This reweighting matches the simulated distribution of the number of inelastic $pp$ collisions per bunch crossing (pile-up) to the one observed in data. The effect of pile-up was modelled by overlaying each hard-scattering event with simulated inelastic $pp$ collisions generated by \PYTHIA[8.186] \cite{Sjostrand:2007gs} using the \NNPDF[2.3lo] set of PDFs \cite{Ball:2012cx} and the A3 tune \cite{ATL-PHYS-PUB-2016-017}. The simulations are further corrected with efficiency scale factors and a smearing of the energy scale of photons, leptons and jets, to better describe the data.
 
Table~\ref{tab:mc_samples_summary} presents a summary of the signal and background samples used in the analysis.

\begin{table}[h!]
\centering
\caption{Details of the MC samples considered in this analysis as GGM signal and SM backgrounds.}
\label{tab:mc_samples_summary}
\resizebox{\textwidth}{!}{
\begin{tabular}{llllrrr}
\toprule
\toprule
 
Process       & Generators                        & PDF sets      & Order \\
\midrule
GGM signals   & \MADGRAPH[5]\,/\,\PYTHIA[8]            & \NNPDF[2.3lo]     & LO  \\ 
{\ttgam}      &  MG5\_aMC{@}NLO\,/\,\PYTHIA[8]      & \NNPDF[2.3lo]     & NLO  \\ 
{\Vgamma}     & \SHERPA[2.2.2]                  & \NNPDF[3.0nnlo]    & 0,1j@NLO + 2,3j@LO \\ 
{\gjet}       & \SHERPA[2.2.2]                  & \NNPDF[3.0nnlo]    & 1,2j@NLO + 3,4j@LO \\ 
{\diphoton}   & \SHERPA[2.2.4]                  & \NNPDF[3.0nnlo]    & 0,1,2,3j@LO \\ 
 
\bottomrule
\bottomrule
\end{tabular}}
\end{table}


\section{Reconstruction of candidates and observables}
\label{sec:objects}

This analysis is performed using the full Run-2 dataset of LHC $pp$ collisions at \COMtev\ collected by the ATLAS
detector between 2015 and 2018, corresponding to an integrated luminosity
of \lumi\ after the application of beam, detector and data quality requirements \cite{DAPR-2018-01}.

The data sample selected by a single-photon trigger with a transverse momentum (\pt) threshold of 140~\gev\
consists of events with at least one photon satisfying the `loose' identification
criteria \cite{TRIG-2018-05}. This trigger is the lowest-threshold unprescaled trigger (considering the complete data-taking period) and is fully efficient for
photons with $\pt>145~\gev$ accepted by the signal selection requirements described in Ref.~\cite{EGAM-2018-01}.

The vertex with the highest sum of the squared transverse momenta of its associated tracks that is reconstructed from at least two good-quality tracks with  $\pt > 0.5$~\gev\ is defined as the primary vertex \cite{ATL-PHYS-PUB-2015-026}. 
After the trigger selection, events are removed from the data sample if they contain jets likely to be produced by beam-induced backgrounds, cosmic rays, or detector noise. 
Photon, lepton, and jet candidates are selected with baseline requirements as described below. Those used to define the different control, validation and signal regions are required to fulfil extra requirements, and are called `signal-region candidates' in the following.

In the offline selection, photon candidates are required to satisfy the
`tight' identification criteria for the lateral and longitudinal shower
shape \cite{EGAM-2018-01}, have $\pt > 25$~\gev\ and $|\eta| < 2.37$, and are removed if they are within the
electromagnetic calorimeter (ECAL) barrel--endcap transition region defined by $1.37<|\eta|<1.52$. An extra
requirement of $\pt>50~\gev$ is imposed on signal-region photons.\footnote{In the final
selection the leading photon is always required to have $\pt>145~\gev$ to be in the maximal efficiency region.}

To reduce the background from jets that can be misidentified as photons, both track and calorimetric isolation requirements
are applied to signal-region photon candidates. The calorimetric isolation energy, {\etiso}, is computed as
the sum of the topological cluster transverse energies \cite{PERF-2014-07} calibrated at the electromagnetic (EM) scale
within a cone of size $\Delta R = 0.4$ around the cluster barycentre. This {\etiso} is required to be less than
$2.45~\gev + 0.022 \pt$, where \pt is that of the photon. The track isolation variable, \ptIsol, is obtained as the scalar sum
of the transverse momenta of good-quality tracks inside a cone of size $\Delta R = 0.2$ around the candidate, and is required to be less than 0.05$\pt$.

Electron candidates are required to have $\pt>10~\gev$ and $|\eta|<2.47$, and to originate from the primary vertex in both the $r$--$z$ and $r$--$\phi$ planes.
A `loose' set of identification criteria are imposed \cite{EGAM-2018-01} and these are based on the characteristics of the EM shower development, the quality of the associated reconstructed track, and the angular proximity of the track to the calorimeter energy deposition.
Signal-region electrons must satisfy `loose' isolation criteria, have $\pt>25~\gev$ and not be within the ECAL barrel--endcap transition region.

Muons are reconstructed by combining compatible track information from the MS and the ID. Muon candidates are required to have $\pt>10~\gev$ and $|\eta|<2.7$, to satisfy the `medium' quality criteria \cite{ATLAS:2020auj} and to originate from the primary vertex in both the $r$--$z$ and $r$--$\phi$ planes.
Signal-region muons must have $\pt>25~\gev$ and satisfy a `loose' isolation requirement.

Jets are reconstructed using the anti-$k_t$ algorithm \cite{Cacciari:2008gp,Fastjet} with a radius parameter $R = 0.4$ and are seeded by the energy in topological clusters of calorimeter cells \cite{PERF-2014-07}. The expected average energy contribution from pile-up interactions is subtracted using a factor that depends on the jet area. Track-based selection requirements are applied to reject jets with $\pt< 120~\gev$ and $|\eta|< 2.4$ that originate from pile-up interactions \cite{PERF-2014-03}.
Except for the \MET\ computation (defined in the following), where a requirement of $|\eta|<4.5$ is applied, jets are kept only if they are in the central region of the detector and have $\pt>20~\gev$.
Signal-region jets must have $\pt>30~\gev$ and $|\eta|<2.5$.

Although jets containing $b$-hadrons, called $b$-jets from now on, are not explicitly used in the analysis selection, they are used in the definition of control regions from which the {\wgamma} and
\ttbargam\ MC normalization is extracted, as is described in \Sec~\ref{sec:selection}. These $b$-jets are selected with the same $\pt$ requirement of jets within the ID acceptance ($|\eta|<2.5$) and identified by the MV2 algorithm, which uses the long lifetime, high decay multiplicity, hard fragmentation and
large mass of $b$-hadrons to distinguish them from light-flavour jets (jets originated from light quarks and gluons)
\cite{ATLAS:2019bwq}. The $b$-tagging algorithm has a nominal efficiency
of 77\% for $b$-jets in simulated \ttbar events and the corresponding probability of misidentifying light-flavour jets is below 1\%.

Due to possible final-state object misidentification, a single object can be reconstructed as more than one object and thus effectively counted multiple times. A procedure to remove these overlaps is applied to preselected objects before the corresponding isolation requirements are imposed. The basic strategy and the order of removal is described in
Refs.~\cite{SUSY-2016-04,ATLAS:2020uiq}.

The missing transverse momentum is computed with an object-based algorithm considering objects passing baseline requirements. Calorimeter energy deposits are matched to high-\pt\ objects in the following order: electrons, photons, jets and muons. Primary-vertex tracks not associated with any such objects are included in the so-called soft term \cite{ATLAS-CONF-2018-023} contribution to \met. The \met\ is computed from the negative vector sum of the transverse momenta of calibrated reconstructed physical objects and the soft term.


\section{Event selection}
\label{sec:selection}


The analysis is designed to compare the event yields observed in three signal regions for strong production (named as SRL, SRM and SRH) with the predictions of the rates of SM processes. 
The SRL region targets the phase space with large mass differences between the gluino and the neutralino, resulting in events characterized by high jet multiplicity and hadronic activity but moderate missing transverse momentum. The SRH region is optimized for the compressed scenarios, near the diagonal in the gluino--neutralino mass plane, giving events with high \MET, higher-\pt\ photons, and lower jet multiplicity and hadronic activity than SRL. Finally, the SRM region is defined for the intermediate phase space between SRL and SRH.

Given the high mass of the gluinos produced in the GGM model-space explored, the total visible transverse momentum is expected to be large. This results in a large value for the variable \HT, defined as the scalar sum of the transverse momenta of all individual signal jets and the leading photon in the final state. 
The selection of signal events includes a requirement on \HT\ as well as on \MET.
In SRL, SRM and SRH, events must contain at least one isolated photon with $\ET$ above $145~\gev$, $300~\gev$ or $400~\gev$, respectively, and zero leptons in order to remove SM $V\gamma$ events where the vector boson decays leptonically. In addition, more than four jets are required in SRL and SRM, while more than two jets are required in SRH.
 
In events characterized by large reconstructed \MET\ without a significant contribution from non-interacting particles or arising from instrumental sources and poorly reconstructed physics objects, the \MET\ vector tends to
be aligned with either the photon or one of the two leading jets. A
selection based on the angular separation between these objects and the \MET\ vector, i.e.\ $\Delta\phi(\text{jet}, \met)$ and  $\Delta\phi(\gam, \met)$, removes most events from these background processes.
 
SUSY signals considered in this analysis are characterized by high-\pt\ multijet
events in a wide region of the parameter space. The sub-leading jets have comparatively larger \pt\ than those in SM background events. Consequently, for signal processes with high-\pt\ jets, \rtfour\ (defined as the ratio of the scalar sum of the \pt\ of the four highest-\pt\ signal-region jets to the scalar sum of the \pt\ of all signal-region jets in the event) takes values less than one, while for SM backgrounds with fewer and softer jets, \rtfour\ is typically closer to unity \cite{PhysRevD.84.055010,SUSY-2016-27}. 
No \rtfour\ selection is applied for SRH because fewer jets are required in this region.

The event selection for all the signal regions is summarized in Table \ref{tab:sr_selection}.

\begin{table}[ht!]
\centering
\caption{Selection for SRL, SRM  and SRH regions.}
\begin{tabular}{lrrr}
\toprule
\toprule
&        SRL    &       SRM     &         SRH \\
\midrule
$N_{\mathrm{photons}}$                        &        $\ge1$ &        $\ge1$ &        $\ge1$ \\ [0.5ex]
$\pt^{\text{leading-}\gamma}$                 &  $>145\ \gev$ &  $>300\ \gev$ &  $>400\ \gev$ \\
$N_{\mathrm{leptons}}$                        &             0 &             0 &             0 \\
$N_{\mathrm{jets}}$                           &       $\ge 5$ &       $\ge 5$ &       $\ge 3$ \\
$\Delta\phi(\text{jet}, \met)$                &        $>0.4$ &        $>0.4$ &        $>0.4$ \\
$\Delta\phi(\gamma, \met)$                    &        $>0.4$ &        $>0.4$ &        $>0.4$ \\
{\met}                                        &  $>250\ \gev$ &  $>300\ \gev$ &  $>600\ \gev$ \\
{\HT}                                         & $>2000\ \gev$ & $>1600\ \gev$ & $>1600\ \gev$ \\
$R_{\mathrm{T}}^4$                            &       $<0.90$ &       $<0.90$ &             - \\
\bottomrule
\bottomrule
\end{tabular}
\label{tab:sr_selection}
\end{table}


\section{Background estimation}
\label{sec:background}


Several SM processes can give final states with real photons and {\met} from the presence of neutrinos. Other SM processes can emulate the targeted topologies if a jet or an electron is misidentified as a photon. The estimation of these backgrounds in the different SRs is therefore essential.

The dominant SM background contributions to the SRs are expected to be from
$W\gamma$ and \ttbargam\ production, followed by prompt photon production with instrumental (fake) \etmiss. These three contributions are determined using MC simulations constrained by the number of data events observed in dedicated control regions through the estimation of normalization factors. The smaller backgrounds, from $W\gamma\gamma$, $Z\gamma$, $Z\gamma\gamma$ and $\gamma\gamma$, are estimated
directly from MC simulation. The backgrounds with misidentified jets or electrons are determined with data-driven techniques as described in the following.
 
Control regions labeled CRW, CRT, and CRQ are used to obtain the MC
normalization for the $W\gamma$, \ttbargam, and QCD \gjet\ events, respectively.
The selection criteria for the CRs associated with the SRs are presented in Table \ref{tab:CR_VR_selection}.
CRs were designed to be orthogonal but still kinematically similar to SRs, and enhanced in the background process of interest, with negligible signal contamination.
Looser {\met} requirements are applied in the CRs to increase the yields. No {\rtfour} requirement is applied for the same reason. An upper bound is placed on {\met} in CRW and CRT to reduce the signal contamination.
 
The selection for CRQ is based on SRL but 
applies a lower \MET\ requirement ($>100~\gev$), a similar \HT\ selection
and an inverted $\Delta\phi(\text{jet}, \met)$ selection to increase the fraction of
\gjet\ in the control sample.

The CRW sample is defined by requiring a photon, a lepton and $100~\gev < \met <  200~\gev$. A $b$-jet veto requirement is applied to reduce the contamination from \ttbargam.
 
The CRT sample is defined by requiring a photon, a lepton, jets and $50~\gev < \met < 200~\gev$. At least two $b$-tagged jets are required in order to increase the sample's purity in {\ttbargam} events.

A further set of event selections define validation regions (VRs) used to check
the results of the background estimation procedure. They were designed to be
kinematically between the signal regions and the control regions, but with one or more criteria
inverted or modified to reduce possible signal contamination. The VRL regions were
designed to be enriched in $W\gamma$ and $\ttbargam$ backgrounds. No $b$-jets requirement is applied,
so contributions from both backgrounds are expected. The four regions (VRL1--4) cover
different parts of the parameter space between the control regions and signal regions by varying the \MET\ and \HT\ requirements.
A signal-region-like VRQ is designed to be orthogonal to the SR only because of a reduced requirement on \MET.
The VRM regions are built to validate the extrapolation of the $\gjet$ background from the CR to the SR.
Two sets of VRMs were specifically designed to select either events with low jet multiplicity and a high-\pt leading photon
(VRM1H and VRM2H) or events with high jet multiplicity and a less energetic photon (VRM1L and VRM2L), in order to validate the background estimation in regions
closer to SRH or SRL respectively. By selecting different ranges in \MET, VRM2L is included in VRM1L, and VRM2H is included in VRM1H.
A summary of the different selection criteria is shown in Table~\ref{tab:CR_VR_selection}.

\begin{table}[ht!]
\centering
\caption{Selection criteria for the control regions and the validation regions used to validate the {\gjet}, $W\gamma$, $\ttbargam$ and $e\to\gamma$ fakes backgrounds.}
\resizebox{\textwidth}{!}{
\begin{tabular}{lcccccccccc}
\toprule
\toprule
Regions & $N_{\mathrm{photons}}$ & $\pt^{\text{leading-}\gamma}$ [\gev] & $N_{\mathrm{leptons}}$ & $N_{\mathrm{jets}}$ & $N_{b\textrm{-jets}}$ & $\Delta\phi(\text{jet}, \MET)$ & $\Delta\phi(\gamma, \MET)$ & \MET[\gev] & \HT\  [\gev] & $R_{\mathrm{T}}^4$  \\
\midrule
CRQ   & $\ge1$ & $>145$ & ~~~~0      & $\ge3$ & -      & $<0.4$ & $>0.4$ & ~~~~$>100$      & ~~~~$>1600$      &  -       \\
CRW   & $\ge1$ & $>145$ & $\ge1$ & $\ge1$ & ~~~~0      & $>0.4$ & -      & $[100,200]$ & ~~~~~~$>400$       &  -       \\
CRT   & $\ge1$ & $>145$ & $\ge1$ & $\ge2$ & $\ge2$ & $>0.4$ & -      & ~~$[50 ,200]$ & ~~~~~~$>400$       &  -       \\
\midrule
VRL1  & $\ge1$ & $>145$ & $\ge1$ & $\ge2$ & -      & $>0.4$ & -      & ~~$[50 ,200]$ & ~~~~~~$>800$       &  -       \\
VRL2  & $\ge1$ & $>145$ & $\ge1$ & $\ge2$ & -      & $>0.4$ & -      & ~~$[50 ,200]$ & ~~~~$>1300$      &  -       \\
VRL3  & $\ge1$ & $>145$ & $\ge1$ & $\ge2$ & -      & $>0.4$ & -      & ~~~~$>200$      & $[600,1600]$ &  -       \\
VRL4  & $\ge1$ & $>145$ & $\ge1$ & $\ge2$ & -      & $<0.4$ & -      & ~~~~$>200$      & ~~~~$>1100$      &  -       \\
\midrule
VRQ   & $\ge1$ & $>145$ & ~~~~0      & $\ge3$ & -      & $>0.4$ & $>0.4$ & $[100,200]$ & ~~~~$>1600$      &  -       \\
VRM1L & $\ge1$ & $>145$ & ~~~~0      & $\ge5$ & -      & $>0.4$ & $>0.4$ & $[100,200]$ & ~~~~$>1600$      &  $<0.90$ \\
VRM2L & $\ge1$ & $>145$ & ~~~~0      & $\ge5$ & -      & $>0.4$ & $>0.4$ & $[150,200]$ & ~~~~$>1600$      &  $<0.90$ \\
VRM1H & $\ge1$ & $>300$ & ~~~~0      & $\ge3$ & -      & $>0.4$ & $>0.4$ & $[100,200]$ & ~~~~$>1600$      &  -       \\
VRM2H & $\ge1$ & $>300$ & ~~~~0      & $\ge3$ & -      & $>0.4$ & $>0.4$ & $[150,200]$ & ~~~~$>1600$      &  -       \\
\midrule
VRE   & $\ge1$ & $>145$ & -      & $\ge1$ & $\ge1$ & $>0.4$ & $<0.4$ & ~~~~$>200$      & $[100,1600]$ &  -       \\
\bottomrule
\bottomrule
 
\end{tabular}
}
\label{tab:CR_VR_selection}
\end{table}

Jets can be misidentified as photons (called `fake photons') if they contain mostly \pizero\ mesons (or any other neutral hadrons) carrying most of the
jet energy and decaying into a pair of collimated photons, resulting in an electromagnetic object 
resembling a single, highly energetic photon.
This background arises primarily from QCD multijets, \wjets\ and semileptonic \ttbar\ events. The `tight' identification criteria applied to photon candidates reduce this background. After applying this selection, the data sample is expected to contain real photons with moderate jet
contamination. As this misidentification rate is not expected to be modelled accurately
in MC simulation, a data-driven sideband counting method \cite{ATLAS:2020uiq} is used. The so-called ABCD
method makes use of the different isolation profiles expected for real photons and misidentified jets \cite{STDM-2010-08}.
Two variables are considered simultaneously in order to include both tracking and calorimetric isolation of the photon candidate, as defined in Section~\ref{sec:objects}. 
The `tight' offline identification is by design tighter than the
photon trigger used to collect the data, so it is expected that some photon
candidates from misidentified jets will fail the `tight' selection but satisfy an intermediate
selection. These photon-like jets, hereinafter called non-tight photons, are defined as those passing the loose identification and satisfying the `tight' selection requirements, except at least one of four selections associated with energy deposits in the EM calorimeter \cite{ATLAS:2020uiq}, chosen to be largely uncorrelated with the isolation variables. In this manner, the use of non-tight photons enhances the `jets faking photons' contribution, as needed by the ABCD method. 
In the identification--isolation plane, the method defines a signal region $A$ consisting of isolated photon candidates that satisfy the `tight' identification, and three control regions, namely $B$, $C $ and $D$, with photon candidates being non-isolated and `tight', isolated and non-tight, and non-isolated and non-tight, respectively.
 
A possible residual correlation between the photon identification and the isolation is estimated using MC simulations, and so is the contamination of the background regions by real photons. These effects are included as part of the computation of the contribution of misidentified jets in all the regions used in the analysis. The systematic uncertainties of the method are evaluated by varying the definition of the non-tight objects, and considering the differences introduced by the residual correlation between the regions.

Significant contamination in the signal regions from SM processes such as $W/Z$\,+\,jets and \ttbar\ events is expected in cases where one high-\pt\ electron is misidentified as a photon.
This background is estimated by weighting the number of electron events observed in an electron control sample by the electron-to-photon fake rate.
These electron control samples come from the same control, validation and signal regions as in the analysis, but the photon kinematic selections are applied to electrons, and then a high-$\pt$ isolated electron is required and signal photons are vetoed.
To estimate the electron-to-photon fake rate, a method based on a sample of \zee\ data events is used \cite{SUSY-2016-04,ATLAS:2020uiq}. Since the $Z$ boson cannot decay directly into an electron and a photon, the
electron--photon events appearing under the $Z$ peak most likely correspond to
misidentified electrons. However, the same applies to other particles decaying
into pairs of electrons. 
Therefore, a background subtraction technique is applied, and this also takes into account the
contamination from random combinatorics background. The electron-to-photon fake factor is then estimated as the ratio of the number of electron--photon pairs to the number of electron--electron pairs found
under the $Z$ peak when fitting the invariant mass distribution.
This fit uses a double-sided
Crystal Ball (DSCB) function (a Gaussian core with asymmetric non-Gaussian power-law tails) to model the $Z$ peak, and a Gaussian distribution to model the small non-resonant backgrounds to \zee\ production.
Only the pairs within a defined invariant mass window are selected to compute the electron-to-photon fake factor. This window is defined as $\pm3\sigma$ around the centre of the peak in the DSCB function, where $\sigma$ is the width of the peak.
Only events with $\etmiss<40~\gev$ are selected,
to avoid electrons from the $W$ decays.

A dedicated validation region (VRE) is designed with the event selection described in Table
\ref{tab:CR_VR_selection}, to validate the accuracy of the corresponding electron-to-photon background
predictions based on the calculated fake factors. The set of requirements mostly selects
$W$($e\nu$)\,+\,jets events, where a boosted $W$ boson (including those coming from top quarks) decays into a
neutrino (giving high {\MET}) and an almost collinear high-{\pt} electron (misidentified as a
photon).

Likelihood fits \cite{Baak:2014wma} are performed assuming i) a background-only hypothesis to estimate the total background in the SRs and VRs; ii) a model-dependent signal plus background hypothesis where the fit is performed in the CRs and SRs simultaneously; and iii) a model-independent signal plus background hypothesis, where both the CRs and SRs are used in the same manner as for the model-dependent signal fit with the number of signal events in the SRs added as a parameter to the fit. This approach constrains the expected background to the yields observed in the data using the CRs and to reduce the systematic uncertainties. Figure \ref{fig:regions_pulls} shows the contributions of the backgrounds in all the different control and validation regions. They are obtained with a `background-only maximum-likelihood fit', constraining the normalizations of the dominant backgrounds and including those estimated using data-driven techniques. The lower panel shows the differences, in standard deviations, between the observed and expected yields \cite{Choudalakis2012}.
Good agreement is found between data and SM background predictions in all the validation regions.
 
\begin{figure}[ht!]
\centering
\includegraphics[width=0.89\textwidth]{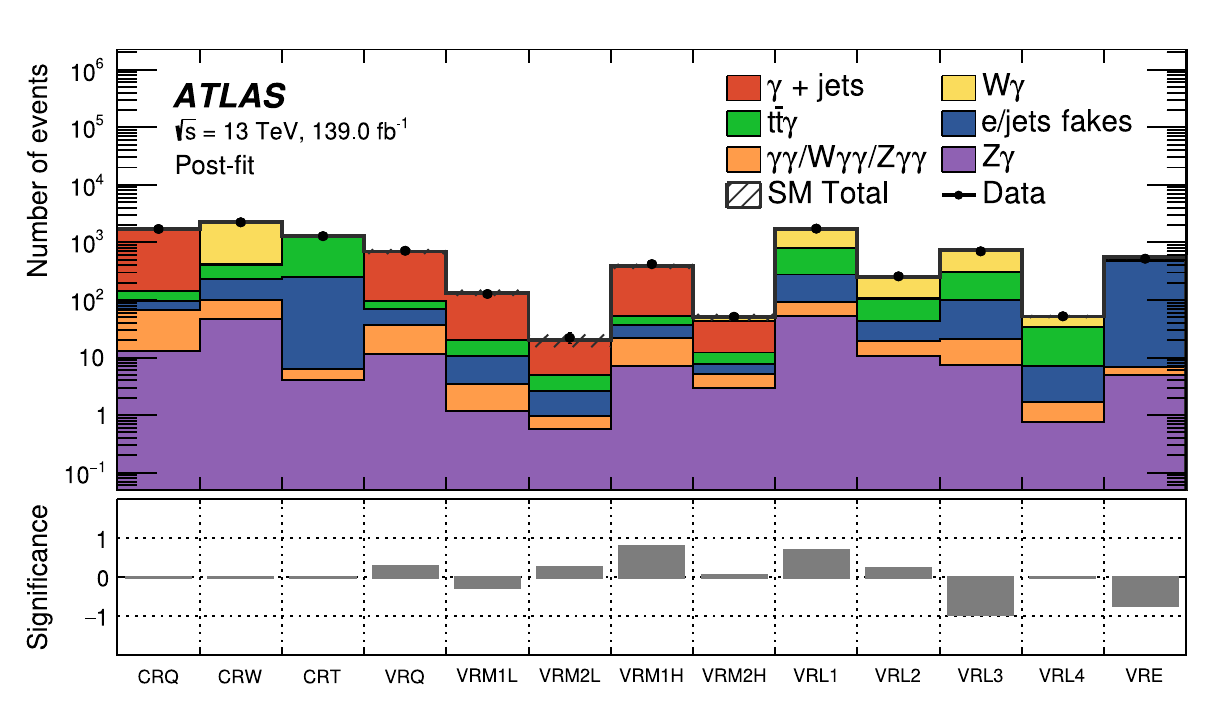}
\caption{The observed and expected (post-fit) yields in the control and validation regions. 
The lower panel shows the difference,
in standard deviations, between the observed and expected yields, considering both the systematic and statistical uncertainties of the
background expectation.}
\label{fig:regions_pulls}
\end{figure}


\section{Systematic uncertainties and yields}
\label{sec:sig_eff}


All background processes estimated either by
making use of MC simulations or by data-driven methods, as well as MC signal predictions,
are affected by systematic uncertainties which mainly originate from two kinds of sources:
experimental and theoretical ones. These systematic uncertainties can impact the
expected event yields in both the control and signal regions.

The uncertainty in the combined 2015--2018 integrated luminosity is 1.7\% \cite{ATLAS-CONF-2019-021}, obtained using the LUCID-2 detector \cite{Avoni_2018} for the primary luminosity measurements. 
The uncertainty on the pile-up reweighting is also considered. 
 
The systematic uncertainties due to the photon identification and isolation efficiencies are estimated
following the prescriptions in Ref.~\cite{EGAM-2018-01}. They are evaluated by varying the correction factors for the photon
selection efficiencies in MC simulation by the corresponding uncertainties. The photon energy scale is determined using samples of $Z\to ee$ events, varying the scale corrections and resolutions upwards and downwards by one standard deviation.

For electrons \cite{EGAM-2018-01} and muons \cite{PERF-2015-10}, similarly
to photons, the uncertainties from the identification efficiency, energy scale and
resolution were determined from $Z \to \ell^{+}\ell^{-}$ and
$W^{\pm} \to \ell^{\pm} \nu$ control samples.
 
For jets, the energy scale and resolution uncertainties are derived following the
procedure described in Ref.~\cite{PERF-2016-04}, where a simplified scheme with 38 parameters is used. 
A set of $b$-tagging uncertainties are also considered, taking an envelope around the nominal jet weight for the selection of different flavour jets \cite{ATLAS:2019bwq}.
 
For {\MET}, the uncertainties associated with all underlying objects from which it is
constructed are propagated through the calculation, and additional uncertainties accounting for the scale and resolution of the soft term \cite{ATLAS-CONF-2018-023} are considered.
For the fake-photon backgrounds ({\jfake} and {\efake}), there are
two different kinds of uncertainties affecting their estimations: the systematic uncertainty from the method used to estimate the fake factors and the
statistical uncertainty of the control sample.

For each of the main simulated background samples, a theoretical uncertainty is assessed by
considering different sources of systematic uncertainty. Each sample
contains several internal weights representing the effect of varying
different parameters of the theory.
The systematic variations considered
for each sample are variations of the renormalization and factorization scales $\muR$ and $\muF$ at generator level, and variations of the PDFs \cite{Butterworth:2015oua} and the strong coupling constant ($\alphas$).
For $\muR$ and $\muF$, three independent nuisance parameters are used, two constructed to keep each of the scales constant while varying the other one, and one as a coherent variation of both scales. 
The PDF uncertainty is taken from an envelope of the nominal PDF (\NNPDF[3.0]) and the variations. Finally, uncertainties associated with $\alphas$ determination and truncation are considered. The PDF and $\alphas$-related uncertainties are added in quadrature.
The total theoretical systematic uncertainty in the signal regions is between 15\% and 30\% depending on the MC sample.
 
The relative impact of each systematic uncertainty on the SM background expectation after the background-only fit applied to the CRs is presented in Table~\ref{tab:syst_rel_impact}. One of the largest experimental systematic uncertainties is related to the jet energy scale and resolution (except for SRH where it is smaller because of the lower jet multiplicity and hadronic activity). Theoretical systematic uncertainties are close to 3\% for SRL and SRM, and are the largest for SRH, reaching the 10\% level.
 
\begin{table}[ht!]
\centering
\caption{Summary of the different systematic uncertainties in the SM background expectation (in \%) for the different SRs after the background-only fit applied to the CRs. The individual uncertainties can be correlated and do not necessarily add in quadrature to equal the total (stat.\ + syst.) uncertainty.}
\begin{tabular}{lccc}
\toprule
\toprule
& SRL [\%] & SRM [\%] & SRH [\%] \\
\midrule
Total (stat.\ + syst.) uncertainty           & 28~~~~~      & 25~~~~~      &  17~~~~~ \\
Statistical uncertainty                     & 20~~~~~      & 15~~~~~      &  12~~~~~ \\
\midrule
Jet energy scale and resolution             & 18~~~~~      & 19~~~~~      & 4.1  \\
$b$-tagging calibration                     & 3.2     & 4.3     & 3.6  \\
Fake photons from jets                      & 2.1     & 2.5     & 2.3  \\
MC theory                                   & 3.6     & 3.1     & 10~~~~~   \\
Fake photons from $e$                       & 1.4     & 1.9     & < 1~~~~~~  \\
Electron/photon energy resolution and scale & 5.5     & 1.1     & 4.1  \\
Muon reconstruction and identification      & 2.6     & 1.8     & < 1~~~~~~  \\
Photon identification and isolation                     & 2.6     & 2.1     & 1.1  \\
Pile-up reweighting                         & < 1~~~~~~     & 1.2     & 1.0  \\
\MET\ soft-term scale and resolution        & < 1~~~~~~     & < 1~~~~~~     & < 1~~~~~~  \\
Luminosity                                  & < 1~~~~~~     & < 1~~~~~~     & < 1~~~~~~  \\
\bottomrule
\bottomrule
\end{tabular}
\label{tab:syst_rel_impact}
\end{table}


\section{Results}
\label{sec:result}

The background-only fit is based on the SRs and CRs listed in Tables~\ref{tab:sr_selection} and \ref{tab:CR_VR_selection} and takes into account all the
systematic uncertainties discussed in Section~\ref{sec:sig_eff}, treated as Gaussian-distributed nuisance parameters. When fitting
the CRs and SRs simultaneously, common normalization factors for each of the {\wgamma},
{\ttbarg} and QCD {\photonjet} events are implemented in order to correctly take
into account the other background contributions. Each experimental uncertainty is treated as fully correlated across the CRs and the corresponding SR, and the physics processes considered. The theoretical systematic uncertainties are treated as correlated across the different regions but uncorrelated across the background samples.

\begin{table}[ht!]
\centering
\caption{Observed events and background estimates (post-fit) in the SRL, SRM and SRH signal regions. The uncertainties in the SM background are both systematic and statistical.}

\begin{tabular}{lrrr}
\toprule
\toprule
& SRL & SRM & SRH \\
\midrule
Observed events & 2 & 0 & 5 \\
\midrule
Expected SM events & $2.67 \pm 0.75$ & $2.55 \pm 0.64$ & $2.55 \pm 0.44$ \\
\midrule
$t\bar{t}\gamma$ & $0.70 \pm 0.18$ & $0.87 \pm 0.18$ & $0.22 \pm 0.05$\\
$W\gamma$ & $0.55 \pm 0.37$ & $0.70 \pm 0.42$ & $1.08 \pm 0.21$\\
$\gamma$ + jets & $0.49 \pm 0.29$ & $0.17 \pm 0.10$ & $0.07 \pm 0.01$\\
$Z(\to\nu\nu)\gamma$ & $0.31 \pm 0.11$ & $0.35 \pm 0.12$ & $0.94 \pm 0.28$\\
$\gamma\gamma / W\gamma\gamma / Z\gamma\gamma$ & $0.23 \pm 0.11$ & $0.25 \pm 0.10$ & $0.08 \pm 0.01$\\
Fake photons from $e$ & $0.22 \pm 0.08$ & $0.04 \pm 0.03$ & $0.06 \pm 0.04$\\
Fake photons from jets & $0.15 \pm 0.09$ & $0.14 \pm 0.09$ & $0.09 \pm 0.07$\\
$Z(\to\ell\ell)\gamma$ & $0.03 \pm 0.03$ & $0.03 \pm 0.01$ & --\\
\bottomrule
\bottomrule
\end{tabular}


\label{tab:fit_result_sr}
\end{table}

The \MET\ distributions observed in the signal regions SRL, SRM and SRH are shown in
Figure~\ref{fig:met_n-1_SRL_SRM_SRH_fr2}, compared to the background predictions. The predicted distributions for selected signals with gluino and neutralino
masses near the expected sensitivity of the analysis are also shown for comparison. In each plot, all the SR selection
requirements are applied except for the one on \MET.

\begin{figure}[!hb]
\begin{center}
\subfloat[SRL]{\includegraphics[width=0.5\textwidth]{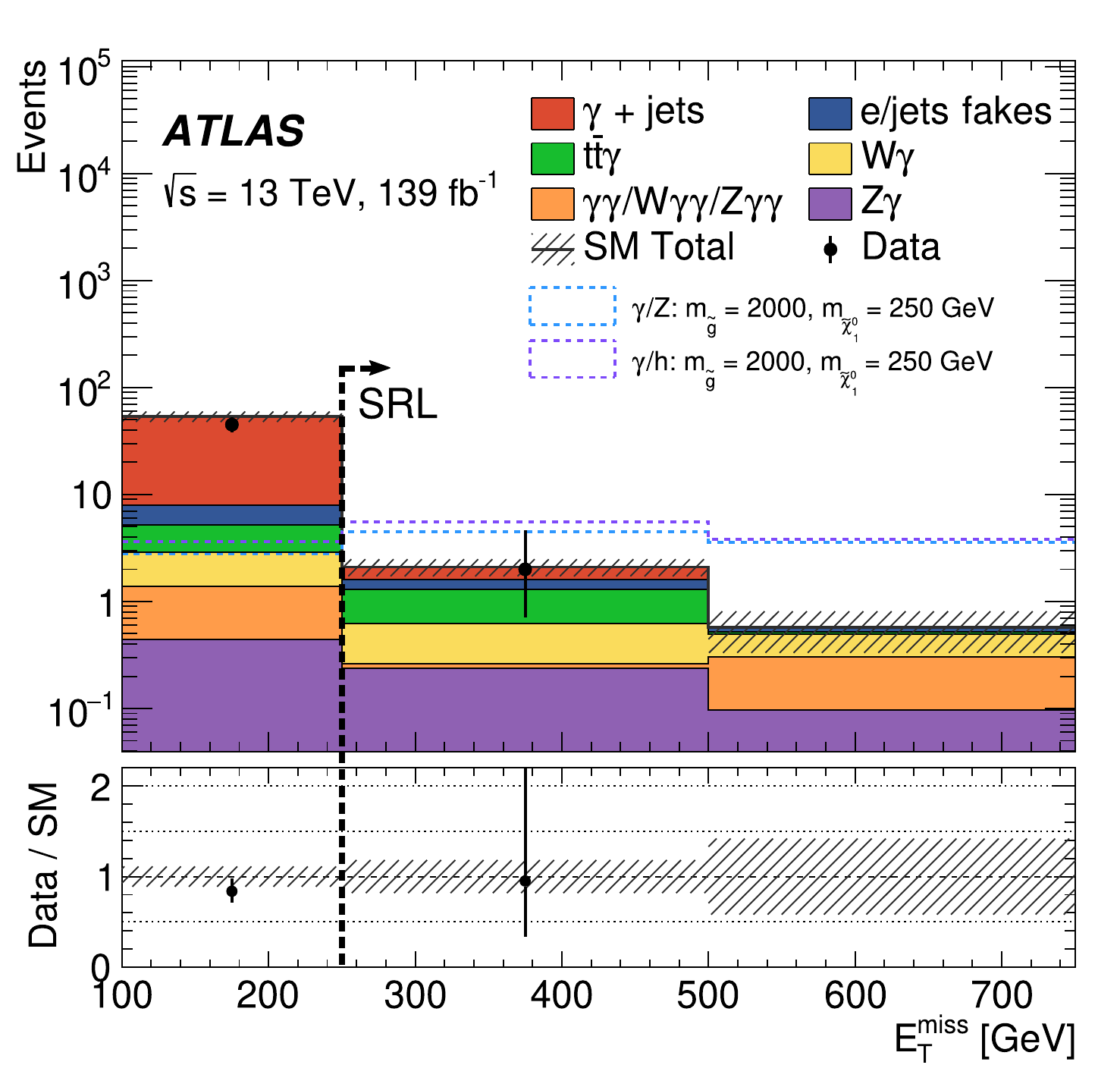}}
\subfloat[SRM]{\includegraphics[width=0.5\textwidth]{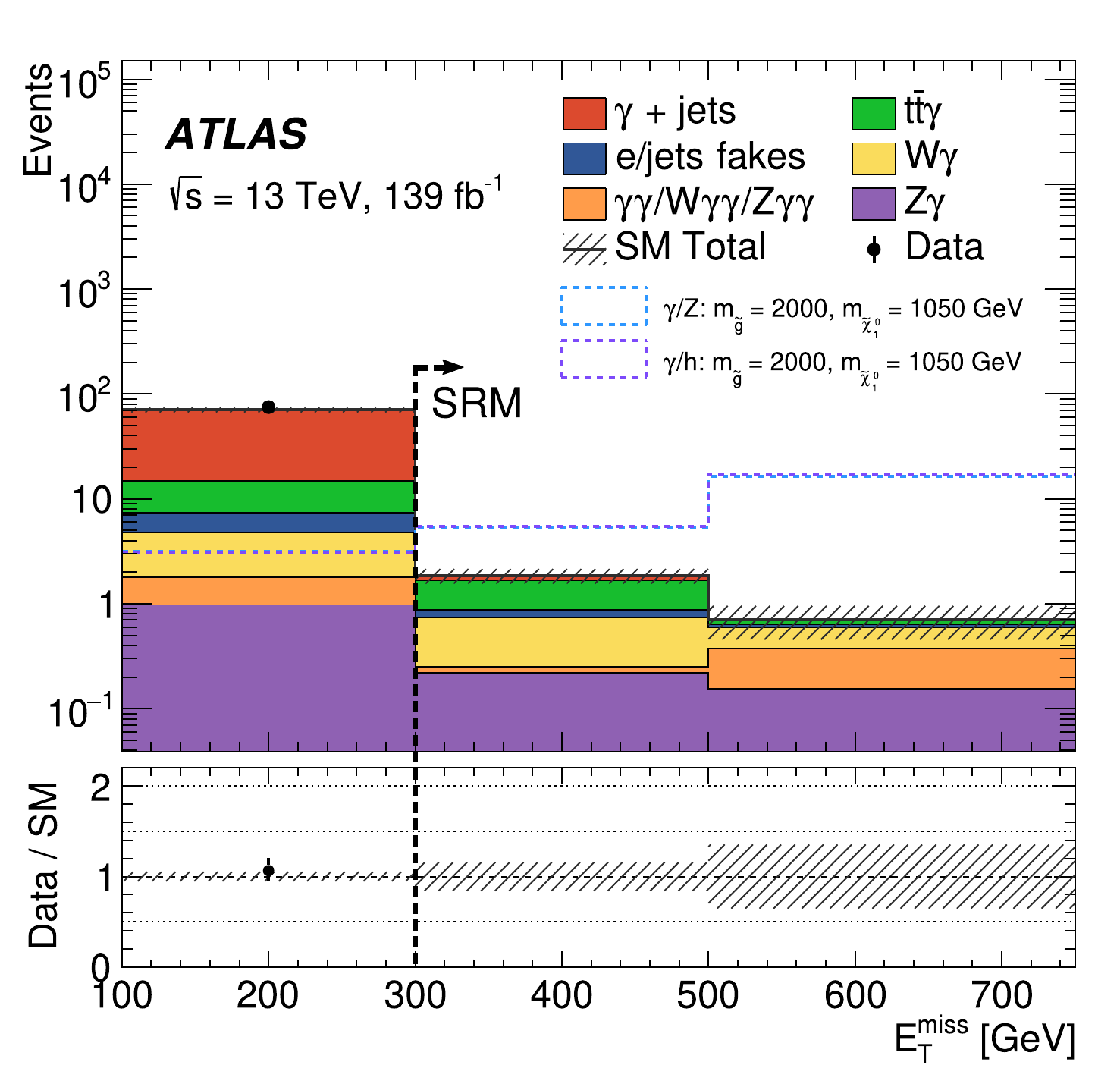}}\\
\subfloat[SRH]{\includegraphics[width=0.5\textwidth]{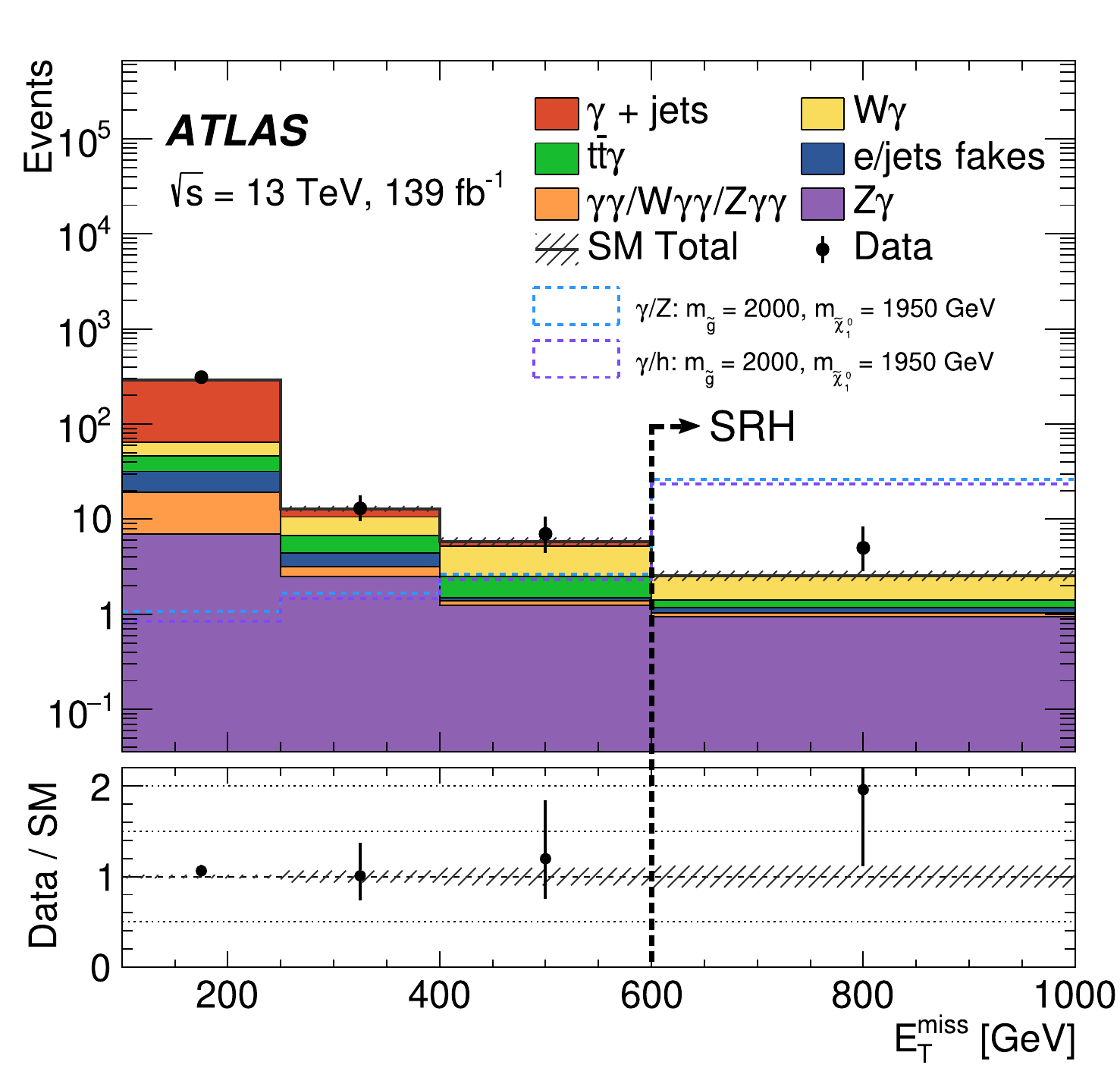}}
\caption{Observed (points with error bars) and expected background (solid histograms) distributions for \MET\ in the signal region (a) SRL, (b) SRM and (c) SRH after the background-only fit applied to the CRs. The predicted signal distributions for the two models with a gluino mass of 2000~\GeV\ and neutralino mass of 250~\GeV\ (SRL), 1050~\GeV\ (SRM) or 1950~\GeV\ (SRH) are also shown for comparison. The uncertainties in the SM background are only statistical.}
\label{fig:met_n-1_SRL_SRM_SRH_fr2}
\end{center}
\end{figure}

The number of data events in each SR and the expected contributions from the different SM backgrounds after the background-only fit applied to the CRs are shown in Table~\ref{tab:fit_result_sr}. Since no significant excess above the SM background is observed in the SRs, these are used to set limits on the number of new-physics events (model-independent limits), and on the GGM signal model parameters described in Section~\ref{sec:intro}.
 
The background-only fit applied to the CRs in previous sections to estimate the background can be extended to include the SRs and perform hypothesis tests, using a profile log-likelihood ratio (LLR) approach \cite{Cowan:2010js}, to assess the compatibility of the observed number of events with the SM, to set limits on the visible cross sections, and to set exclusion limits on specific SUSY models.
 
The model-independent limits on the number of events from
non-SM processes in each SR are listed in Table~\ref{tab:model_indep_ul}, together with the discovery $p$-value ($p_{0}$), defined as the probability of observing at least the observed event yield when assuming that no signal is present, and the corresponding Gaussian significance $Z$. Also
shown is the 95\% confidence level (CL) upper limit on the visible cross section $\sigma \times A \times \epsilon$, obtained by normalizing the upper limit on the number of signal events to the integrated luminosity, where $\sigma$ is the production cross section for a beyond-the-SM (BSM) signal, $A$ is the acceptance (fraction of events with objects passing all the kinematic selections at particle level) and $\epsilon$ is the efficiency (fraction of those events that would be observed after the detector reconstruction).
For SRL and SRM, $p_{0}$ is capped at 0.5 because the predictions exceed the data. For SRH, the discovery $p$-value is 0.09, so these observations are compatible with the SM-only hypothesis.
The number of observed events and the background expectation in each SR is used to set a 95\% CL upper limit on the number of events from any BSM physics scenario \cite{Read:2002hq}. 
The most stringent observed limit is from SRM, where visible cross sections greater than 0.022~fb are excluded. 
 
\begin{table}[!h]
\centering
\caption{
Summary of the  model-independent limits results, with
the 95\% CL upper limits on the visible cross section
($\langle\epsilon\sigma\rangle_{\mathrm{obs}}^{95}$) and on the number of
signal events ($S_{\mathrm{obs}}^{95}$ ).
The values of $\langle\epsilon\sigma\rangle_{\mathrm{exp}}^{95}$ and $S_{\mathrm{exp}}^{95}$
are the expected 95\% CL upper limits on the visible cross section and the number of
signal events, given the expected number (and $\pm 1\sigma$ excursions of
the expectation) of background events. Finally, the $p$-value ($p_0$) and the corresponding Gaussian significance
$Z$ for the three signal regions are shown.} 
 
\begin{tabular}{lccccc}
\toprule
\toprule
Signal Region & $\langle\epsilon{\sigma}\rangle_{\mathrm{obs}}^{95}$ [fb]  & $\langle\epsilon{\sigma}\rangle_{\mathrm{exp}}^{95}$ [fb] & $S_{\mathrm{obs}}^{95}$  & $S_{\mathrm{exp}}^{95}$ & $p_{0}$ ($Z$)\\
\midrule
SRL  &  0.034   &   $0.034^{+0.016}_{-0.009}$   &   4.7   &   $4.7^{+2.2}_{-1.2}$    &    0.50 (0.00)  \\ [0.5ex]
SRM  &  0.022   &   $0.033^{+0.013}_{-0.008}$   &   3~~~      &   $4.6^{+1.8}_{-1.1}$    &    0.50 (0.00)  \\ [0.5ex]
SRH  &  0.054   &   $0.035^{+0.014}_{-0.010}$   &   7.6   &   $4.8^{+1.9}_{-1.4}$    &    0.09 (1.32)  \\
\bottomrule
\bottomrule
\end{tabular}
\label{tab:model_indep_ul}
\end{table}

The exclusion limit for a specific SUSY signal model is based on the profile LLR test statistic, and it is obtained from a simultaneous fit to the contributions from SM processes and the targeted model in a given signal region and its associated background control regions, which are all by design statistically independent. These one-sided limits are set at the 95\% CL using the CL$_\mathrm{s}$ prescription \cite{Read:2002hq}. 
The observed exclusion limit is calculated with
signal yields corresponding to the nominal cross section $\pm 1 \sigma$ of the SUSY
theoretical uncertainty. The combined exclusion limits are shown in Figure~\ref{fig:limit_plot_combined}, for each of the two signal models considered. These are obtained with pseudo-data experiments, and using the signal region with the best expected sensitivity at each point.
The black dashed line corresponds to the expected limits at 95\% CL, with the light
(yellow) bands indicating the $\pm 1 \sigma$ exclusions due to experimental and background-theory uncertainties. The observed limits are indicated by medium
(red) curves, the solid contour represents the nominal limit, and the
dotted lines are obtained by varying the signal cross section by the theoretical
scale and PDF uncertainties.
The discontinuity in the transition from high to moderate NLSP masses is due to the number of events observed in SRH (SRM) being higher (lower) than the expected value.

\begin{figure}[ht!]
\centering
\subfloat[$\gamma/Z$ model]{\includegraphics[width=0.5\textwidth]{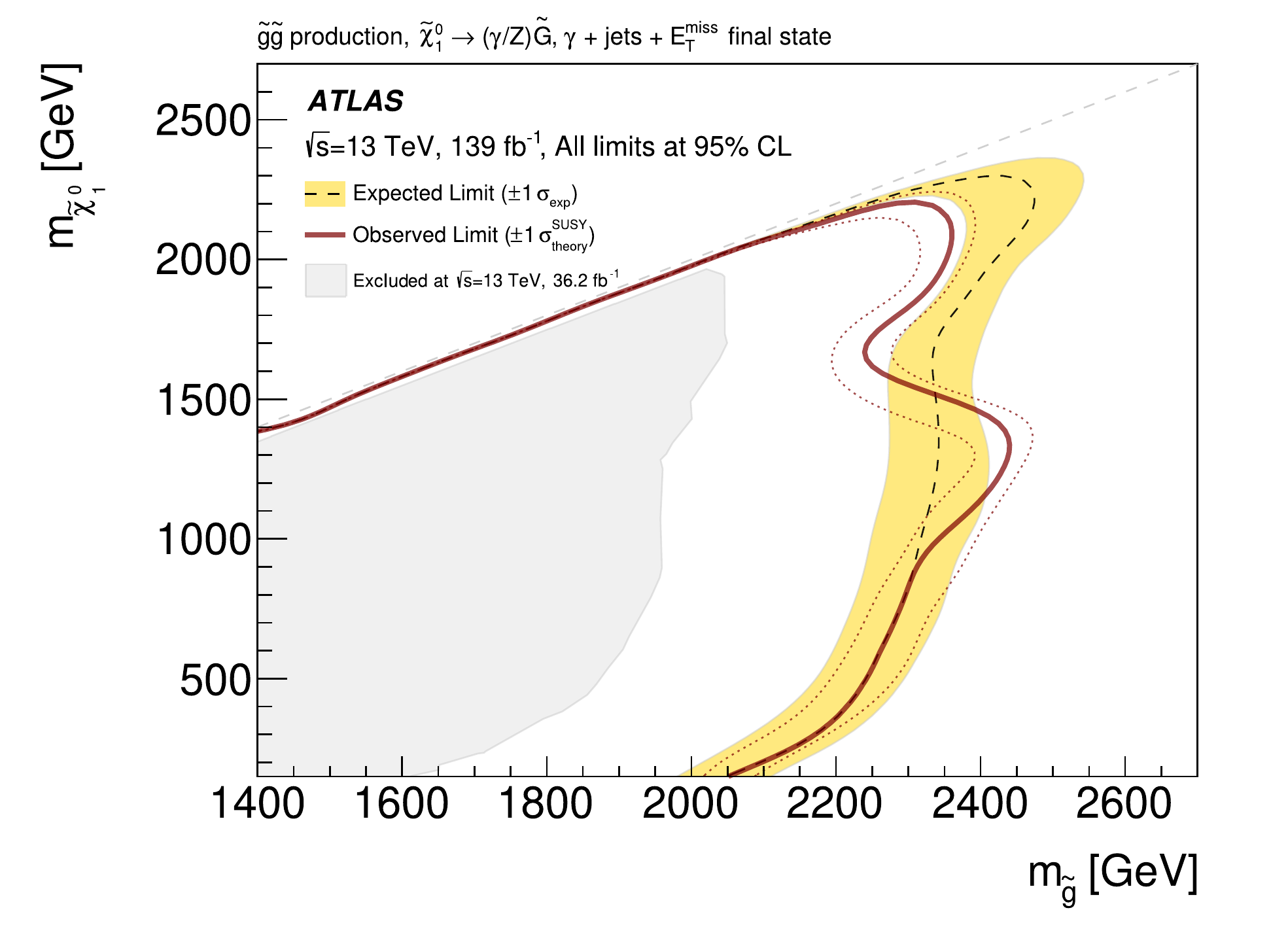}}
\subfloat[$\gamma/h$ model]{\includegraphics[width=0.5\textwidth]{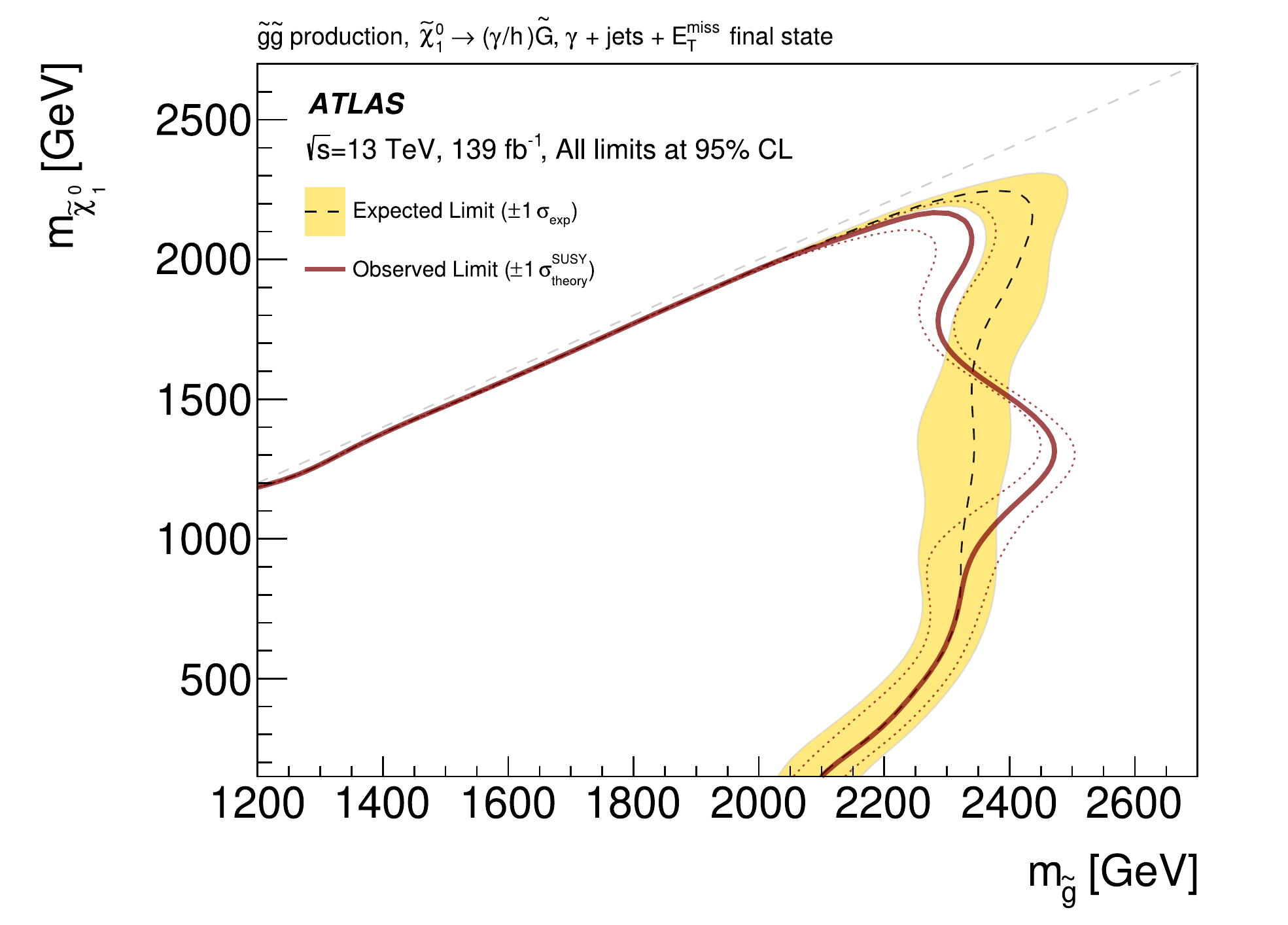}}
 
\caption{Observed and expected exclusion limits in the gluino--neutralino mass plane at 95\% CL for the full Run-2 dataset corresponding to an integrated luminosity of 139~\ifb, for the (a) $\gamma/Z$ and (b) $\gamma/h$ signal models. They are obtained by combining limits from the signal region with the best expected sensitivity at each point. The dashed (black) line corresponds to the expected limits at 95\% CL, with the light (yellow) band indicating the $\pm 1 \sigma$ excursions due to experimental and background-theory uncertainties. The observed limits are indicated by medium (red) curves: the solid contour represents the nominal limit, and the dotted lines are obtained by varying the signal cross section by the theoretical scale and PDF uncertainties.}
\label{fig:limit_plot_combined}
\end{figure}

For the $\gamma/Z$ signal model, the lower limits on the gluino mass in this paper are between 200 and 400~\gev\ higher than those obtained in the previous search \cite{SUSY-2016-27}. For the $\gamma/h$ signal model, the previous search \cite{SUSY-2014-01} was performed in Run-1 data with a slightly different mass-plane coverage and set a lower limit around 1.2~\tev\ for the gluino mass. In the present study, the lower limits on the gluino mass in this model are almost 1~\tev\ higher.
For both models, the most stringent lower limit on the gluino mass is set at 2.4~\tev\ for a neutralino mass of 1.3--1.4~\tev. Furthermore, an overall lower limit on the gluino mass of 2.2~\tev\ is obtained for all neutralino masses except those below 150~\gev, which are regions where the analysis is expected to have low signal acceptance, due to trigger constrains and large \met\ requirements in the SR.


\FloatBarrier

\section{Conclusions}
\label{sec:conclusion}

Based on proton--proton collision data at $\sqrt{s} = 13~\tev$ corresponding to an integrated luminosity of 139~\ifb\ recorded by the ATLAS
detector at the LHC in Run 2, a search has been performed for the experimental signature of at least one isolated photon with high transverse momentum, jets and a large amount of missing transverse momentum. Three signal regions are defined, one with a prediction of $2.67 \pm 0.75$ background events and 2 events observed, another with $2.55 \pm 0.64$ background events predicted and no events observed, and the third with $2.55 \pm 0.44$ background events predicted and 5 events observed.
These results are compatible with no significant excess of events over the SM background expectation. Model-dependent 95\% CL upper limits are set on possible contributions from new physics in a GGM scenario with a NLSP neutralino that is a mixture of higgsino and bino. Pair-produced gluinos with masses up to 2200~\gev\ are excluded for most of the NLSP masses investigated, giving the most stringent limits obtained by ATLAS. Model-independent 95\% CL upper limits are set on the visible cross section for contributions from new physics for each of the defined signal regions. The most stringent observed limit is from SRM, where visible cross sections greater than 0.022~fb are excluded.


\section*{Acknowledgements}


We thank CERN for the very successful operation of the LHC, as well as the
support staff from our institutions without whom ATLAS could not be
operated efficiently.
 
We acknowledge the support of
ANPCyT, Argentina;
YerPhI, Armenia;
ARC, Australia;
BMWFW and FWF, Austria;
ANAS, Azerbaijan;
CNPq and FAPESP, Brazil;
NSERC, NRC and CFI, Canada;
CERN;
ANID, Chile;
CAS, MOST and NSFC, China;
Minciencias, Colombia;
MEYS CR, Czech Republic;
DNRF and DNSRC, Denmark;
IN2P3-CNRS and CEA-DRF/IRFU, France;
SRNSFG, Georgia;
BMBF, HGF and MPG, Germany;
GSRI, Greece;
RGC and Hong Kong SAR, China;
ISF and Benoziyo Center, Israel;
INFN, Italy;
MEXT and JSPS, Japan;
CNRST, Morocco;
NWO, Netherlands;
RCN, Norway;
MEiN, Poland;
FCT, Portugal;
MNE/IFA, Romania;
MESTD, Serbia;
MSSR, Slovakia;
ARRS and MIZ\v{S}, Slovenia;
DSI/NRF, South Africa;
MICINN, Spain;
SRC and Wallenberg Foundation, Sweden;
SERI, SNSF and Cantons of Bern and Geneva, Switzerland;
MOST, Taiwan;
TENMAK, T\"urkiye;
STFC, United Kingdom;
DOE and NSF, United States of America.
In addition, individual groups and members have received support from
BCKDF, CANARIE, Compute Canada and CRC, Canada;
PRIMUS 21/SCI/017 and UNCE SCI/013, Czech Republic;
COST, ERC, ERDF, Horizon 2020 and Marie Sk{\l}odowska-Curie Actions, European Union;
Investissements d'Avenir Labex, Investissements d'Avenir Idex and ANR, France;
DFG and AvH Foundation, Germany;
Herakleitos, Thales and Aristeia programmes co-financed by EU-ESF and the Greek NSRF, Greece;
BSF-NSF and MINERVA, Israel;
Norwegian Financial Mechanism 2014-2021, Norway;
NCN and NAWA, Poland;
La Caixa Banking Foundation, CERCA Programme Generalitat de Catalunya and PROMETEO and GenT Programmes Generalitat Valenciana, Spain;
G\"{o}ran Gustafssons Stiftelse, Sweden;
The Royal Society and Leverhulme Trust, United Kingdom.
 
The crucial computing support from all WLCG partners is acknowledged gratefully, in particular from CERN, the ATLAS Tier-1 facilities at TRIUMF (Canada), NDGF (Denmark, Norway, Sweden), CC-IN2P3 (France), KIT/GridKA (Germany), INFN-CNAF (Italy), NL-T1 (Netherlands), PIC (Spain), ASGC (Taiwan), RAL (UK) and BNL (USA), the Tier-2 facilities worldwide and large non-WLCG resource providers. Major contributors of computing resources are listed in Ref.~\cite{ATL-SOFT-PUB-2021-003}.


\clearpage
 
\printbibliography

@article{Viel_2005,
	doi = {10.1103/physrevd.71.063534},
	year = "2005",
	volume = {71},
	number = {6},
      eprint = "astro-ph/0501562",
      archivePrefix = "arXiv",
      primaryClass = "astro-ph",
	author = "Viel, M. and Lesgourgues, J. and Haehnelt, M. G. and Matarrese, S. and Riotto, A.",
	title = "{Constraining warm dark matter candidates including sterile neutrinos and light gravitinos with {WMAP} and the Lyman-$\ensuremath{\alpha}$ forest}",
	journal = {Phys. Rev. D}
}

@article{Kamada2014,
    author = "Kamada, A. and Shirasaki, M. and Yoshida, N.",
    title = "{Weighing the light gravitino mass with weak lensing surveys}",
    eprint = "1311.4323",
    archivePrefix = "arXiv",
    primaryClass = "hep-ph",
    doi = "10.1007/JHEP06(2014)162",
    journal = "JHEP",
    volume = "2014",
    pages = "162",
    year = "2014"
}

@article{Martin:1997ns,
    author = "Martin, S. P.",
    title = "{A Supersymmetry primer}",
    eprint = "9709356",
    archivePrefix = "arXiv",
    primaryClass = "hep-ex",
    doi = "10.1142/9789812839657_0001",
    journal = "Adv. Ser. Direct. High Energy Phys.",
    volume = "18",
    pages = "1--98",
    year = "1998"
}

@article{Shih2012,
  title = "{General neutralino NLSPs at the early LHC}",
  author = "{J. Ruderman and D. Shih}",
  journal = "JHEP",
  volume = "08",
  pages = "159",
  numpages = "21",
  year = "2012",
  eprint = "1103.6083",
  archivePrefix = "arXiv",
  primaryClass = "hep-ex",
  doi = {10.1007/JHEP08(2012)159},
}

@article{Choudalakis2012,
  title = "{Plotting the differences between data and expectation}",
  author = "{G. Choudalakis and D. Casadei}",
  journal = "Eur. Phys. J. Plus",
  volume = "127",
  pages = "25",
  numpages = "10",
  year = "2012",
  eprint = "1111.2062",
  archivePrefix = "arXiv",
  primaryClass = "physics.data-an",
  doi = {10.1140/epjp/i2012-12025-y},
}

@article{PhysRevD.84.055010,
  title = "{Event-shape selection cuts for supersymmetry searches at the LHC with 7 TeV energy}",
  author = "{M. Guchait and D. Sengupta}",
  journal = "Phys. Rev. D",
  volume = "84",
  pages = "055010",
  numpages = "6",
  year = "2011",
  month = "Sep",
  eprint = "1102.4785",
  archivePrefix = "arXiv",
  primaryClass = "hep-ex",
  doi = {10.1103/PhysRevD.84.055010},
}

@article{ATLAS:2020uiq,
    author = "{ATLAS Collaboration}",
    title = "{Search for dark matter in association with an energetic photon in $pp$ collisions at $\sqrt{s}$ = 13 TeV with the ATLAS detector}",
    eprint = "2011.05259",
    archivePrefix = "arXiv",
    primaryClass = "hep-ex",
    reportNumber = "CERN-EP-2020-178",
    doi = "10.1007/JHEP02(2021)226",
    journal = "JHEP",
    volume = "02",
    pages = "226",
    year = "2021"
}

@article{ATLAS:2019bwq,
    author = "{ATLAS Collaboration}",
    title = "{ATLAS b-jet identification performance and efficiency measurement with $t{\bar{t}}$ events in pp collisions at $\sqrt{s}=13$ TeV}",
    eprint = "1907.05120",
    archivePrefix = "arXiv",
    primaryClass = "hep-ex",
    reportNumber = "CERN-EP-2019-132",
    doi = "10.1140/epjc/s10052-019-7450-8",
    journal = "Eur. Phys. J. C",
    volume = "79",
    number = "11",
    pages = "970",
    year = "2019"
}

@article{ATLAS:2020auj,
    author = "{ATLAS Collaboration}",
    title = "{Muon reconstruction and identification efficiency in ATLAS using the full Run 2 $pp$ collision data set at $\sqrt{s}=13$ TeV}",
    journal = "Eur. Phys. J. C",
    volume = "81",
    pages = "578",
    doi={10.1140/epjc/s10052-021-09233-2},
    eprint = "2012.00578",
    archivePrefix = "arXiv",
    primaryClass = "hep-ex",
    reportNumber = "CERN-EP-2020-199",
    month = "12",
    year = "2020"
}

@article{Avoni_2018,
	doi = "10.1088/1748-0221/13/07/p07017",
%	url = {https://doi.org/10.1088/1748-0221/13/07/p07017},
	year = "2018",
	month = "jul",
	volume = "13",
	number = "07",
	pages = "P07017",
% author = "{ATLAS Collaboration}",
	author = "{G. Avoni et al.}",
	title = "{The new {LUCID}-2 detector for luminosity measurement and monitoring in {ATLAS}}",
	journal = "JINST",
}

@article{Geant4,
      author         = "Agostinelli, S. and others",
      title          = "{\textsc{Geant4} -- a simulation toolkit}",
      collaboration  = "GEANT4",
      journal        = "Nucl. Instrum. Meth. A",
      volume         = "506",
      year           = "2003",
      pages          = "250-303",
      doi            = "10.1016/S0168-9002(03)01368-8",
      reportNumber   = "SLAC-PUB-9350, FERMILAB-PUB-03-339",
      SLACcitation   = "%%CITATION = NUIMA,A506,250;%%"
}

@article{Djouadi2007426,
title = "SuSpect: A Fortran code for the Supersymmetric and Higgs particle spectrum in the MSSM",
journal = "Comput. Phys. Commun. ",
volume = "176",
number = "6",
pages = "426 - 455",
year = "2007",
note = "",
doi = "10.1016/j.cpc.2006.11.009",
author = "Abdelhak Djouadi and Jean-Loïc Kneur and Gilbert Moultaka"
}

@article{Djouadi:1997yw,
      author         = "Djouadi, A. and Kalinowski, J. and Spira, M.",
      title          = "{HDECAY: a program for Higgs boson decays in the Standard
                        Model and its supersymmetric extension}",
      journal        = "Comput. Phys. Commun.",
      volume         = "108",
      pages          = "56-74",
      doi            = "10.1016/S0010-4655(97)00123-9",
      year           = "1998",
      eprint         = "hep-ph/9704448",
      archivePrefix  = "arXiv",
      primaryClass   = "hep-ph",
      reportNumber   = "DESY-97-079, IFT-96-29, PM-97-04",
      SLACcitation   = "%%CITATION = HEP-PH/9704448;%%",
}

@article{MUHLLEITNER200546,
title = "SDECAY: a Fortran code for the decays of the supersymmetric particles in the MSSM",
journal = "Comput. Phys. Commun.",
volume = "168",
number = "1",
pages = "46-70",
year = "2005",
eprint         = "hep-ph/0311167",
archivePrefix  = "arXiv",
primaryClass   = "hep-ph",
doi = "https://doi.org/10.1016/j.cpc.2005.01.012",
author = "M. Mühlleitner and A. Djouadi and Y. Mambrini"
}

@article{Djouadi:2006bz,
      author         = "Djouadi, A. and Muhlleitner, M.M. and Spira, M.",
      title          = "{Decays of supersymmetric particles: The Program SUSY-HIT
                        (SUspect-SdecaY-Hdecay-InTerface)}",
      journal        = "Acta Phys. Polon. B",
      volume         = "38",
      pages          = "635-644",
      year           = "2007",
      eprint         = "hep-ph/0609292",
      archivePrefix  = "arXiv",
      primaryClass   = "hep-ph",
      reportNumber   = "CERN-PH-TH-2006-200",
      SLACcitation   = "%%CITATION = HEP-PH/0609292;%%",
}

@article{Borschensky:2014cia,
      author         = "Borschensky, Christoph and Kr{\"a}mer, Michael and Kulesza,
                        Anna and Mangano, Michelangelo and Padhi, Sanjay and
                        Plehn, Tilman and Portell, Xavier",
      title          = "{Squark and gluino production cross sections in pp
                        collisions at \(\sqrt{s} = 13, 14, 33\) and \(100\,\text{TeV}\)}",
      journal        = "Eur. Phys. J. C",
      volume         = "74",
      year           = "2014",
      pages          = "3174",
      doi            = "10.1140/epjc/s10052-014-3174-y",
      eprint         = "1407.5066",
      archivePrefix  = "arXiv",
      primaryClass   = "hep-ph",
      reportNumber   = "MS-TP-14-25, CERN-PH-TH-2014-137, TTK-14-13",
      SLACcitation   = "%%CITATION = ARXIV:1407.5066;%%"
}

@article{Farrar:1978xj,
      author         = "Farrar, Glennys R. and Fayet, Pierre",
      title          = "{Phenomenology of the production, decay, and detection of
                        new hadronic states associated with supersymmetry}",
      journal        = "Phys. Lett. B",
      volume         = "76",
      year           = "1978",
      pages          = "575",
      doi            = "10.1016/0370-2693(78)90858-4",
      reportNumber   = "CALT-68-648",
      SLACcitation   = "%%CITATION = PHLTA,B76,575;%%"
}

@article{Ellis:1983ew,
      author         = "Ellis, John and Hagelin, J.S. and Nanopoulos, Dimitri V. and Olive, Keith A. and Srednicki, M.",
      title          = "{Supersymmetric relics from the big bang}",
      booktitle      = "{IN *BATAVIA 1984, PROCEEDINGS, INNER SPACE/OUTER SPACE*,
                        458-459., In *Srednicki, M.A. (ed.): Particle physics and
                        cosmology* 223-246}",
      journal        = "Nucl. Phys. B",
      volume         = "238",
      year           = "1984",
      pages          = "453",
      doi            = "10.1016/0550-3213(84)90461-9",
      reportNumber   = "SLAC-PUB-3171",
      SLACcitation   = "%%CITATION = NUPHA,B238,453;%%"
}

@article{Dine:1981gu,
      author         = "Dine, Michael and Fischler, Willy",
      title          = "{A phenomenological model of particle physics based on supersymmetry}",
      journal        = "Phys. Lett. B",
      volume         = "110",
      year           = "1982",
      pages          = "227",
      doi            = "10.1016/0370-2693(82)91241-2",
      reportNumber   = "PRINT-82-0029 (IAS,PRINCETON)",
      SLACcitation   = "%%CITATION = PHLTA,B110,227;%%"
}

@article{AlvarezGaume:1981wy,
      author         = "Alvarez-Gaumé, Luis and Claudson, Mark and Wise, Mark B.",
      title          = "{Low-energy supersymmetry}",
      journal        = "Nucl. Phys. B",
      volume         = "207",
      year           = "1982",
      pages          = "96",
      doi            = "10.1016/0550-3213(82)90138-9",
      reportNumber   = "HUTP-81-A063",
      SLACcitation   = "%%CITATION = NUPHA,B207,96;%%"
}

@article{Nappi:1982hm,
      author         = "Nappi, Chiara R. and Ovrut, Burt A.",
      title          = "{Supersymmetric extension of the SU(3) x SU(2) x U(1) model}",
      journal        = "Phys. Lett. B",
      volume         = "113",
      year           = "1982",
      pages          = "175",
      doi            = "10.1016/0370-2693(82)90418-X",
      reportNumber   = "PRINT-82-0093 (IAS,PRINCETON)",
      SLACcitation   = "%%CITATION = PHLTA,B113,175;%%"
}

@article{Cheung:2007es,
      author         = "Cheung, Clifford and Fitzpatrick, A. Liam and Shih, David",
      title          = "{(Extra)ordinary gauge mediation}",
      journal        = "JHEP",
      volume         = "07",
      year           = "2008",
      pages          = "054",
      doi            = "10.1088/1126-6708/2008/07/054",
      eprint         = "0710.3585",
      archivePrefix  = "arXiv",
      primaryClass   = "hep-ph",
      SLACcitation   = "%%CITATION = ARXIV:0710.3585;%%"
}

@article{Meade:2008wd,
      author         = "Meade, Patrick and Seiberg, Nathan and Shih, David",
      title          = "{General Gauge Mediation}",
      booktitle      = "{30 years of mathematical methods in high energy physics.
                        Proceedings, International Conference in honor of
                        Professor Tohru Eguchi's 60th birthday, Kyoto, Japan,
                        March 17-19, 2008}",
      journal        = "Prog. Theor. Phys. Suppl.",
      volume         = "177",
      year           = "2009",
      pages          = "143",
      doi            = "10.1143/PTPS.177.143",
      eprint         = "0801.3278",
      archivePrefix  = "arXiv",
      primaryClass   = "hep-ph",
      SLACcitation   = "%%CITATION = ARXIV:0801.3278;%%"
}

@article{Beenakker:1996ch,
      author         = "Beenakker, W. and H{\"o}pker, R. and Spira, M. and Zerwas, P.M.",
      title          = "{Squark and gluino production at hadron colliders}",
      journal        = "Nucl. Phys. B",
      volume         = "492",
      pages          = "51",
      doi            = "10.1016/S0550-3213(97)00084-9",
      year           = "1997",
      eprint         = "hep-ph/9610490",
      archivePrefix  = "arXiv",
      reportNumber   = "DESY-96-150, CERN-TH-96-215",
      SLACcitation   = "%%CITATION = HEP-PH/9610490;%%",
}

@article{Kulesza:2008jb,
      author         = "Kulesza, A. and Motyka, L.",
      title          = "{Threshold Resummation for Squark-Antisquark and Gluino-Pair Production at the LHC}",
      journal        = "Phys. Rev. Lett.",
      volume         = "102",
      pages          = "111802",
      doi            = "10.1103/PhysRevLett.102.111802",
      year           = "2009",
      eprint         = "0807.2405",
      archivePrefix  = "arXiv",
      primaryClass   = "hep-ph",
      reportNumber   = "DESY-08-096",
      SLACcitation   = "%%CITATION = ARXIV:0807.2405;%%",
}

@article{Kulesza:2009kq,
      author         = "Kulesza, A. and Motyka, L.",
      title          = "{Soft gluon resummation for the production of gluino-gluino and squark-antisquark pairs at the LHC}",
      journal        = "Phys. Rev. D",
      volume         = "80",
      pages          = "095004",
      doi            = "10.1103/PhysRevD.80.095004",
      year           = "2009",
      eprint         = "0905.4749",
      archivePrefix  = "arXiv",
      primaryClass   = "hep-ph",
      reportNumber   = "DESY-09-078, PITHA-09-11",
      SLACcitation   = "%%CITATION = ARXIV:0905.4749;%%",
}

@article{Beenakker:2009ha,
      author         = "Beenakker, Wim and Brensing, Silja and Kr{\"a}mer, Michael and Kulesza, Anna and Laenen, Eric and Niessen, Irene",
      title          = "{Soft-gluon resummation for squark and gluino hadroproduction}",
      journal        = "JHEP",
      volume         = "12",
      pages          = "041",
      doi            = "10.1088/1126-6708/2009/12/041",
      year           = "2009",
      eprint         = "0909.4418",
      archivePrefix  = "arXiv",
      primaryClass   = "hep-ph",
      reportNumber   = "ITP-UU-09-33, NIKHEF-2009-015",
      SLACcitation   = "%%CITATION = ARXIV:0909.4418;%%",
}

@article{Beenakker:2011fu,
      author         = "Beenakker, W. and others",
      title          = "{Squark and gluino hadroproduction}",
      journal        = "Int. J. Mod. Phys. A",
      volume         = "26",
      pages          = "2637",
      doi            = "10.1142/S0217751X11053560",
      year           = "2011",
      eprint         = "1105.1110",
      archivePrefix  = "arXiv",
      primaryClass   = "hep-ph",
      reportNumber   = "DESY-11-060, ITFA-11-09, ITP-UU-11-13, NIKHEF-2011-011,
                        TTK-11-12",
      SLACcitation   = "%%CITATION = ARXIV:1105.1110;%%",
}

@article{Baak:2014wma,
      author         = "Baak, M. and Besjes, G.J. and C{\^o}t{\'e}, D. and Koutsman, A. and Lorenz, J. and Short, D.",
      title          = "{HistFitter software framework for statistical data analysis}",
      journal        = "Eur. Phys. J. C",
      volume         = "75",
      year           = "2015",
      pages          = "153",
      doi            = "10.1140/epjc/s10052-015-3327-7",
      eprint         = "1410.1280",
      archivePrefix  = "arXiv",
      primaryClass   = "hep-ex",
      SLACcitation   = "%%CITATION = ARXIV:1410.1280;%%",
}

@Article{Cacciari:2008gp,
     author    = "Cacciari, Matteo and Salam, Gavin P. and Soyez, Gregory",
     title     = "{The anti-\(k_{t}\) jet clustering algorithm}",
     journal   = "JHEP",
     volume    = "04",
     year      = "2008",
     pages     = "063",
     eprint    = "0802.1189",
     archivePrefix = "arXiv",
     primaryClass  =  "hep-ph",
     doi       = "10.1088/1126-6708/2008/04/063",
     SLACcitation  = "%%CITATION = 0802.1189;%%"
}

@Article{Fastjet,
      author         = "Cacciari, Matteo and Salam, Gavin P. and Soyez, Gregory",
      title          = "{FastJet user manual}",
      journal        = "Eur. Phys. J. C",
      volume         = "72",
      year           = "2012",
      pages          = "1896",
      doi            = "10.1140/epjc/s10052-012-1896-2",
      eprint         = "1111.6097",
      archivePrefix  = "arXiv",
      primaryClass   = "hep-ph",
      reportNumber   = "CERN-PH-TH-2011-297",
      SLACcitation   = "%%CITATION = ARXIV:1111.6097;%%"
}

@Article{Butterworth:2015oua,
      author         = "Butterworth, Jon and others",
      title          = "{PDF4LHC recommendations for LHC Run II}",
      journal        = "J. Phys. G",
      volume         = "43",
      year           = "2016",
      pages          = "023001",
      doi            = "10.1088/0954-3899/43/2/023001",
      eprint         = "1510.03865",
      archivePrefix  = "arXiv",
      primaryClass   = "hep-ph",
      reportNumber   = "OUTP-15-17P, SMU-HEP-15-12, TIF-UNIMI-2015-14,
                        LCTS-2015-27, CERN-PH-TH-2015-249",
      SLACcitation   = "%%CITATION = ARXIV:1510.03865;%%"
}

@article{Ball:2012cx,
      author         = "{NNPDF Collaboration} and Ball, Richard D. and others",
      title          = "{Parton distributions with LHC data}",
      journal        = "Nucl. Phys. B",
      volume         = "867",
      year           = "2013",
      pages          = "244",
      doi            = "10.1016/j.nuclphysb.2012.10.003",
      eprint         = "1207.1303",
      archivePrefix  = "arXiv",
      primaryClass   = "hep-ph",
      reportNumber   = "EDINBURGH-2012-08, IFUM-FT-997, FR-PHENO-2012-014,
                        RWTH-TTK-12-25, CERN-PH-TH-2012-037, SFB-CPP-12-47\,
                        --CERN-PH-TH-2012-037",
      SLACcitation   = "%%CITATION = ARXIV:1207.1303;%%"
}

@article{Ball:2014uwa,
      author         = "{The NNPDF Collaboration} and Ball, Richard D. and others",
      title          = "{Parton distributions for the LHC run II}",
      journal        = "JHEP",
      volume         = "04",
      year           = "2015",
      pages          = "040",
      doi            = "10.1007/JHEP04(2015)040",
      eprint         = "1410.8849",
      archivePrefix  = "arXiv",
      primaryClass   = "hep-ph",
      reportNumber   = "EDINBURGH-2014-15, IFUM-1034-FT, CERN-PH-TH-2013-253,
                        OUTP-14-11P, CAVENDISH-HEP-14-11",
      SLACcitation   = "%%CITATION = ARXIV:1410.8849;%%"
}

@Article{Sjostrand:2007gs,
	author    = "Sj{\"o}strand, T. and Mrenna, S. and Skands, P.",
	title     = "{A brief introduction to PYTHIA 8.1}",
	journal   = "Comput. Phys. Commun.",
	volume    = "178",
	year      = "2008",
	pages     = "852-867",
	eprint    = "0710.3820",
	archivePrefix = "arXiv",
	primaryClass  =  "hep-ph",
	doi       = "10.1016/j.cpc.2008.01.036",
	SLACcitation  = "%%CITATION = 0710.3820;%%"
}

@article{Sjostrand:2014zea,
      author         = "Sj{\"o}strand, Torbj{\"o}rn and Ask, Stefan and Christiansen,
                        Jesper R. and Corke, Richard and Desai, Nishita and Ilten,
                        Philip and Mrenna, Stephen and Prestel, Stefan and
                        Rasmussen, Christine O. and Skands, Peter Z.",
      title          = "{An introduction to PYTHIA 8.2}",
      journal        = "Comput. Phys. Commun.",
      volume         = "191",
      year           = "2015",
      pages          = "159",
      doi            = "10.1016/j.cpc.2015.01.024",
      eprint         = "1410.3012",
      archivePrefix  = "arXiv",
      primaryClass   = "hep-ph",
      reportNumber   = "LU-TP-14-36, MCNET-14-22, CERN-PH-TH-2014-190,
                        FERMILAB-PUB-14-316-CD, DESY-14-178, SLAC-PUB-16122,
                        --FERMILAB-PUB-14-316-CD",
      SLACcitation   = "%%CITATION = ARXIV:1410.3012;%%"
}

@Article{Alwall:2014hca,
      author         = "Alwall, J. and Frederix, R. and Frixione, S. and Hirschi,
                        V. and Maltoni, F. and Mattelaer, O. and Shao, H. -S. and
                        Stelzer, T. and Torrielli, P. and Zaro, M.",
      title          = "{The automated computation of tree-level and
                        next-to-leading order differential cross sections, and
                        their matching to parton shower simulations}",
      journal        = "JHEP",
      volume         = "07",
      year           = "2014",
      pages          = "079",
      doi            = "10.1007/JHEP07(2014)079",
      eprint         = "1405.0301",
      archivePrefix  = "arXiv",
      primaryClass   = "hep-ph",
      reportNumber   = "CERN-PH-TH-2014-064, CP3-14-18, LPN14-066, MCNET-14-09,
                        ZU-TH-14-14",
      SLACcitation   = "%%CITATION = ARXIV:1405.0301;%%"
}

@Article{Hoeche:2009rj,
      author         = "H{\"o}che, Stefan and Krauss, Frank and Schumann, Steffen and Siegert, Frank",
      title          = "{QCD matrix elements and truncated showers}",
      journal        = "JHEP",
      volume         = "05",
      pages          = "053",
      doi            = "10.1088/1126-6708/2009/05/053",
      year           = "2009",
      eprint         = "0903.1219",
      archivePrefix  = "arXiv",
      primaryClass   = "hep-ph",
}

@Article{Gleisberg:2008fv,
      author         = "Gleisberg, Tanju and H{\"o}che, Stefan",
      title          = "{Comix, a new matrix element generator}",
      journal        = "JHEP",
      volume         = "12",
      pages          = "039",
      doi            = "10.1088/1126-6708/2008/12/039",
      year           = "2008",
      eprint         = "0808.3674",
      archivePrefix  = "arXiv",
      primaryClass   = "hep-ph",
}

@Article{Schumann:2007mg,
      author         = "Schumann, Steffen and Krauss, Frank",
      title          = "{A parton shower algorithm based on Catani--Seymour dipole factorisation}",
      journal        = "JHEP",
      volume         = "03",
      pages          = "038",
      doi            = "10.1088/1126-6708/2008/03/038",
      year           = "2008",
      eprint         = "0709.1027",
      archivePrefix  = "arXiv",
      primaryClass   = "hep-ph",
}

@article{Hoeche:2012yf,
      author         = "H{\"o}che, Stefan and Krauss, Frank and Sch{\"o}nherr, Marek and
                        Siegert, Frank",
      title          = "{QCD matrix elements + parton showers. The NLO case}",
      journal        = "JHEP",
      volume         = "04",
      year           = "2013",
      pages          = "027",
      doi            = "10.1007/JHEP04(2013)027",
      eprint         = "1207.5030",
      archivePrefix  = "arXiv",
      primaryClass   = "hep-ph",
      reportNumber   = "SLAC-PUB-15191, IPPP-12-52, DCPT-12-104, LPN12-081,
                        FR-PHENO-2012-017, MCNET-12-09, --FR-PHENO-2012-017",
      SLACcitation   = "%%CITATION = ARXIV:1207.5030;%%"
}

@article{Bothmann:2019yzt,
      author = "Bothmann, Enrico and others",
      title = "{Event generation with Sherpa 2.2}",
      journal = "SciPost Phys.",
      volume = "7",
      year = "2019",
      number = "3",
      pages = "034",
      doi = "10.21468/SciPostPhys.7.3.034",
      reportNumber = "FERMILAB-PUB-19-218-T, SLAC-PUB-17433, IPPP/19/42, MCNET-19-11",
      eprint = "1905.09127",
      archivePrefix = "arXiv",
      primaryClass = "hep-ph",
}

@article{Hoeche:2011fd,
      author         = "H{\"o}che, Stefan and Krauss, Frank and Sch{\"o}nherr, Marek and
                        Siegert, Frank",
      title          = "{A critical appraisal of NLO+PS matching methods}",
      journal        = "JHEP",
      volume         = "09",
      year           = "2012",
      pages          = "049",
      doi            = "10.1007/JHEP09(2012)049",
      eprint         = "1111.1220",
      archivePrefix  = "arXiv",
      primaryClass   = "hep-ph",
      reportNumber   = "SLAC-PUB-14661, IPPP-11-67, DCPT-11-134, LPN11-58,
                        FR-PHENO-2011-019, MCNET-11-24",
      SLACcitation   = "%%CITATION = ARXIV:1111.1220;%%"
}

@article{Catani:2001cc,
      author         = "Catani, S. and Krauss, F. and Webber, B. R. and Kuhn, R.",
      title          = "{QCD Matrix Elements + Parton Showers}",
      journal        = "JHEP",
      volume         = "11",
      year           = "2001",
      pages          = "063",
      doi            = "10.1088/1126-6708/2001/11/063",
      eprint         = "hep-ph/0109231",
      archivePrefix  = "arXiv",
      reportNumber   = "CERN-TH-2000-367, CAVENDISH-HEP-00-03",
      SLACcitation   = "%%CITATION = HEP-PH/0109231;%%"
}

@article{Cascioli:2011va,
      author         = "Cascioli, Fabio and Maierh{\"o}fer, Philipp and Pozzorini, Stefano",
      title          = "{Scattering Amplitudes with Open Loops}",
      journal        = "Phys. Rev. Lett.",
      volume         = "108",
      year           = "2012",
      pages          = "111601",
      doi            = "10.1103/PhysRevLett.108.111601",
      eprint         = "1111.5206",
      archivePrefix  = "arXiv",
      primaryClass   = "hep-ph",
      reportNumber   = "ZU-TH-23-11, LPN11-66",
      SLACcitation   = "%%CITATION = ARXIV:1111.5206;%%"
}

@article{Buccioni:2019sur,
    author = "Buccioni, Federico and Lang, Jean-Nicolas and Lindert, Jonas M. and Maierh{\"o}fer, Philipp and Pozzorini, Stefano and Zhang, Hantian and Zoller, Max F.",
    title = "{OpenLoops 2}",
    eprint = "1907.13071",
    archivePrefix = "arXiv",
    primaryClass = "hep-ph",
    reportNumber = "IPPP/19/62, FR-PHENO-2019-12, PSI-PR-19-15, ZU-TH 37/19",
    doi = "10.1140/epjc/s10052-019-7306-2",
    journal = "Eur. Phys. J. C",
    volume = "79",
    number = "10",
    pages = "866",
    year = "2019"
}

@article{Denner:2016kdg,
      author         = "Denner, Ansgar and Dittmaier, Stefan and Hofer, Lars",
      title          = "{\textsc{Collier}: A fortran-based complex one-loop library in
                        extended regularizations}",
      journal        = "Comput. Phys. Commun.",
      volume         = "212",
      year           = "2017",
      pages          = "220-238",
      doi            = "10.1016/j.cpc.2016.10.013",
      eprint         = "1604.06792",
      archivePrefix  = "arXiv",
      primaryClass   = "hep-ph",
      reportNumber   = "FR-PHENO-2016-003, ICCUB-16-016",
      SLACcitation   = "%%CITATION = ARXIV:1604.06792;%%"
}

@Article{Read:2002hq,
    Author = {Read, Alexander L.},
    Journal = {J. Phys. G},
    Pages = {2693},
    doi = "10.1088/0954-3899/28/10/313",
    Title = "{Presentation of search results: the \(CL_S\) technique}",
    Volume = {28},
    Year = {2002}
}

@Article{Cowan:2010js,
   Author = {Cowan, Glen and Cranmer, Kyle and Gross, Eilam and Vitells, Ofer},
   Title = {{Asymptotic formulae for likelihood-based tests of new physics}},
   Journal = {Eur. Phys. J. C},
   Volume = {71},
   Year = {2011},
   Pages = {1554},
   doi = "10.1140/epjc/s10052-011-1554-0",
   Eprint = "1007.1727",
   Archiveprefix = {arXiv},
   Primaryclass = {physics.data-an},
   related = "Cowan:2010js-err",
   relatedstring  = "Erratum:",
}

@Article{PERF-2007-01,
    author         = "{ATLAS Collaboration}",
    title          = "{The ATLAS Experiment at the CERN Large Hadron Collider}",
    journal        = "JINST",
    volume         = "3",
    year           = "2008",
    pages          = "S08003",
    doi            = "10.1088/1748-0221/3/08/S08003",
    primaryClass   = "hep-ex",
}

@Article{SOFT-2010-01,
    author         = "{ATLAS Collaboration}",
    title          = "{The ATLAS Simulation Infrastructure}",
    journal        = "Eur. Phys. J. C",
    volume         = "70",
    year           = "2010",
    pages          = "823",
    doi            = "10.1140/epjc/s10052-010-1429-9",
    eprint         = "1005.4568",
    archivePrefix  = "arXiv",
    primaryClass   = "physics.ins-det",
}

@Article{STDM-2010-08,
    author         = "{ATLAS Collaboration}",
    title          = "{Measurement of the inclusive isolated prompt photon cross section in \(pp\) collisions at \(\sqrt{s} = 7\,\text{TeV}\) with the ATLAS detector}",
    journal        = "Phys. Rev. D",
    volume         = "83",
    year           = "2011",
    pages          = "052005",
    doi            = "10.1103/PhysRevD.83.052005",
    reportNumber   = "CERN-PH-EP-2010-068",
    eprint         = "1012.4389",
    archivePrefix  = "arXiv",
    primaryClass   = "hep-ex",
}

@Article{PERF-2014-03,
    author         = "{ATLAS Collaboration}",
    title          = "{Performance of pile-up mitigation techniques for jets in \(pp\) collisions at \(\sqrt{s} = 8\,\text{TeV}\) using the ATLAS detector}",
    journal        = "Eur. Phys. J. C",
    volume         = "76",
    year           = "2016",
    pages          = "581",
    doi            = "10.1140/epjc/s10052-016-4395-z",
    reportNumber   = "CERN-PH-EP-2015-206",
    eprint         = "1510.03823",
    archivePrefix  = "arXiv",
    primaryClass   = "hep-ex",
}

@Article{PERF-2014-07,
    author         = "{ATLAS Collaboration}",
    title          = "{Topological cell clustering in the ATLAS calorimeters and its performance in LHC Run~1}",
    journal        = "Eur. Phys. J. C",
    volume         = "77",
    year           = "2017",
    pages          = "490",
    doi            = "10.1140/epjc/s10052-017-5004-5",
    reportNumber   = "CERN-PH-EP-2015-304",
    eprint         = "1603.02934",
    archivePrefix  = "arXiv",
    primaryClass   = "hep-ex",
}

@Article{SUSY-2014-01,
    author         = "{ATLAS Collaboration}",
    title          = "{Search for photonic signatures of gauge-mediated supersymmetry in \(8\,\text{TeV}\) \(pp\) collisions with the ATLAS detector}",
    journal        = "Phys. Rev. D",
    volume         = "92",
    year           = "2015",
    pages          = "072001",
    doi            = "10.1103/PhysRevD.92.072001",
    reportNumber   = "CERN-PH-EP-2015-168",
    eprint         = "1507.05493",
    archivePrefix  = "arXiv",
    primaryClass   = "hep-ex",
}

@Article{PERF-2015-10,
    author         = "{ATLAS Collaboration}",
    title          = "{Muon reconstruction performance of the ATLAS detector in proton--proton collision data at \(\sqrt{s} = 13\,\text{TeV}\)}",
    journal        = "Eur. Phys. J. C",
    volume         = "76",
    year           = "2016",
    pages          = "292",
    doi            = "10.1140/epjc/s10052-016-4120-y",
    reportNumber   = "CERN-EP-2016-033",
    eprint         = "1603.05598",
    archivePrefix  = "arXiv",
    primaryClass   = "hep-ex",
}

@Article{PERF-2016-04,
    author         = "{ATLAS Collaboration}",
    title          = "{Jet energy scale measurements and their systematic uncertainties in proton--proton collisions at \(\sqrt{s} = 13\,\text{TeV}\) with the ATLAS detector}",
    journal        = "Phys. Rev. D",
    volume         = "96",
    year           = "2017",
    pages          = "072002",
    doi            = "10.1103/PhysRevD.96.072002",
    reportNumber   = "CERN-EP-2017-038",
    eprint         = "1703.09665",
    archivePrefix  = "arXiv",
    primaryClass   = "hep-ex",
}

@Article{SUSY-2016-04,
    author         = "{ATLAS Collaboration}",
    title          = "{Search for supersymmetry in a final state containing two photons and missing transverse momentum in \(\sqrt{s} = 13\,\text{TeV}\) \(pp\) collisions at the LHC using the ATLAS detector}",
    journal        = "Eur. Phys. J. C",
    volume         = "76",
    year           = "2016",
    pages          = "517",
    doi            = "10.1140/epjc/s10052-016-4344-x",
    reportNumber   = "CERN-EP-2016-146",
    eprint         = "1606.09150",
    archivePrefix  = "arXiv",
    primaryClass   = "hep-ex",
}

@Article{SUSY-2016-27,
    author         = "{ATLAS Collaboration}",
    title          = "{Search for photonic signatures of gauge-mediated supersymmetry in \(13\,\text{TeV}\) \(pp\) collisions with the ATLAS detector}",
    journal        = "Phys. Rev. D",
    volume         = "97",
    year           = "2018",
    pages          = "092006",
    doi            = "10.1103/PhysRevD.97.092006",
    reportNumber   = "CERN-EP-2017-323",
    eprint         = "1802.03158",
    archivePrefix  = "arXiv",
    primaryClass   = "hep-ex",
}

@Article{TRIG-2016-01,
    author         = "{ATLAS Collaboration}",
    title          = "{Performance of the ATLAS trigger system in 2015}",
    journal        = "Eur. Phys. J. C",
    volume         = "77",
    year           = "2017",
    pages          = "317",
    doi            = "10.1140/epjc/s10052-017-4852-3",
    reportNumber   = "CERN-EP-2016-241",
    eprint         = "1611.09661",
    archivePrefix  = "arXiv",
    primaryClass   = "hep-ex",
}

@Article{DAPR-2018-01,
    author         = "{ATLAS Collaboration}",
    title          = "{ATLAS data quality operations and performance for 2015--2018 data-taking}",
    journal        = "JINST",
    volume         = "15",
    year           = "2020",
    pages          = "P04003",
    doi            = "10.1088/1748-0221/15/04/P04003",
    reportNumber   = "CERN-EP-2019-207",
    eprint         = "1911.04632",
    archivePrefix  = "arXiv",
    primaryClass   = "physics.ins-det",
}

@Article{EGAM-2018-01,
    author         = "{ATLAS Collaboration}",
    title          = "{Electron and photon performance measurements with the ATLAS detector using the 2015--2017 LHC proton--proton collision data}",
    journal        = "JINST",
    volume         = "14",
    year           = "2019",
    pages          = "P12006",
    doi            = "10.1088/1748-0221/14/12/P12006",
    reportNumber   = "CERN-EP-2019-145",
    eprint         = "1908.00005",
    archivePrefix  = "arXiv",
    primaryClass   = "hep-ex",
}

@Article{PIX-2018-001,
    author         = "Abbott, B. and others",
    title          = "{Production and integration of the ATLAS Insertable B-Layer}",
    journal        = "JINST",
    volume         = "13",
    year           = "2018",
    pages          = "T05008",
    doi            = "10.1088/1748-0221/13/05/T05008",
    eprint         = "1803.00844",
    archivePrefix  = "arXiv",
    primaryClass   = "physics.ins-det",
}

@Article{TRIG-2018-05,
    author         = "{ATLAS Collaboration}",
    title          = "{Performance of electron and photon triggers in ATLAS during LHC Run~2}",
    journal        = "Eur. Phys. J. C",
    volume         = "80",
    year           = "2020",
    pages          = "47",
    doi            = "10.1140/epjc/s10052-019-7500-2",
    reportNumber   = "CERN-EP-2019-169",
    eprint         = "1909.00761",
    archivePrefix  = "arXiv",
    primaryClass   = "hep-ex",
}

@Booklet{ATL-SOFT-PUB-2021-001,
    author         = "{ATLAS Collaboration}",
    title          = "{The ATLAS Collaboration Software and Firmware}",
    howpublished   = "{ATL-SOFT-PUB-2021-001}",
    url            = "https://cds.cern.ch/record/2767187",
    year           = "2021",
}

@Booklet{ATL-SOFT-PUB-2021-003,
    author         = "{ATLAS Collaboration}",
    title          = "{ATLAS Computing Acknowledgements}",
    howpublished   = "{ATL-SOFT-PUB-2021-003}",
    url            = "https://cds.cern.ch/record/2776662",
    year           = "2021",
}

@Report{ATLAS-TDR-19,
    author         = "{ATLAS Collaboration}",
    title          = "{ATLAS Insertable B-Layer: Technical Design Report}",
    type           = "ATLAS-TDR-19; CERN-LHCC-2010-013",
    year           = "2010",
    url            = "https://cds.cern.ch/record/1291633",
    related        = "ATLAS-TDR-19-addm",
    relatedstring  = "Addendum:",
}

@Article{CMS-SUS-17-011,
    author         = "{CMS Collaboration}",
    title          = "{Search for supersymmetry in final states with photons and missing transverse momentum in proton--proton collisions at \(13\,\text{TeV}\)}",
    journal        = "JHEP",
    volume         = "06",
    year           = "2019",
    pages          = "143",
    doi            = "10.1007/JHEP06(2019)143",
    reportNumber   = "CERN-EP-2019-040",
    eprint         = "1903.07070",
    archivePrefix  = "arXiv",
    primaryClass   = "hep-ex",
}

@Article{CMS-SUS-18-002,
    author         = "{CMS Collaboration}",
    title          = "{Search for supersymmetry in events with a photon, jets, \(b\)-jets, and missing transverse momentum in proton--proton collisions at \(13\,\text{TeV}\)}",
    journal        = "Eur. Phys. J. C",
    volume         = "79",
    year           = "2019",
    pages          = "444",
    doi            = "10.1140/epjc/s10052-019-6926-x",
    reportNumber   = "CERN-EP-2018-353",
    eprint         = "1901.06726",
    archivePrefix  = "arXiv",
    primaryClass   = "hep-ex",
}

@Article{CMS-SUS-18-005,
    author         = "{CMS Collaboration}",
    title          = "{Combined search for supersymmetry with photons in proton--proton collisions at \(\sqrt{s} = 13\,\text{TeV}\)}",
    journal        = "Phys. Lett. B",
    volume         = "801",
    year           = "2020",
    pages          = "135183",
    doi            = "10.1016/j.physletb.2019.135183",
    reportNumber   = "CERN-EP-2019-114",
    eprint         = "1907.00857",
    archivePrefix  = "arXiv",
    primaryClass   = "hep-ex",
}

@Booklet{ATLAS-CONF-2018-023,
    author         = "{ATLAS Collaboration}",
    title          = "{\(E_{\text{T}}^{\text{miss}}\) performance in the ATLAS detector using 2015--2016 LHC \(pp\) collisions}",
    howpublished   = "{ATLAS-CONF-2018-023}",
    url            = "https://cds.cern.ch/record/2625233",
    year           = "2018",
}

@Booklet{ATLAS-CONF-2019-021,
    author         = "{ATLAS Collaboration}",
    title          = "{Luminosity determination in \(pp\) collisions at \(\sqrt{s} = 13\,\text{TeV}\) using the ATLAS detector at the LHC}",
    howpublished   = "{ATLAS-CONF-2019-021}",
    url            = "https://cds.cern.ch/record/2677054",
    year           = "2019",
}

@Booklet{ATL-PHYS-PUB-2010-013,
    author         = "{ATLAS Collaboration}",
    title          = "{The simulation principle and performance of the ATLAS fast calorimeter simulation FastCaloSim}",
    howpublished   = "{ATL-PHYS-PUB-2010-013}",
    url            = "https://cds.cern.ch/record/1300517",
    year           = "2010",
}

@Booklet{ATL-PHYS-PUB-2014-021,
    author         = "{ATLAS Collaboration}",
    title          = "{ATLAS Pythia~8 tunes to \(7~\text{TeV}\) data}",
    howpublished   = "{ATL-PHYS-PUB-2014-021}",
    url            = "https://cds.cern.ch/record/1966419",
    year           = "2014",
}

@Booklet{ATL-SOFT-PUB-2014-001,
    author         = "{ATLAS Collaboration}",
    title          = "{Performance of the Fast ATLAS Tracking Simulation (FATRAS) and the ATLAS Fast Calorimeter Simulation (FastCaloSim) with single particles}",
    howpublished   = "{ATL-SOFT-PUB-2014-001}",
    url            = "https://cds.cern.ch/record/1669341",
    year           = "2014",
}

@Booklet{ATL-PHYS-PUB-2015-026,
    author         = "{ATLAS Collaboration}",
    title          = "{Vertex Reconstruction Performance of the ATLAS Detector at \(\sqrt{s} = 13~\text{TeV}\)}",
    howpublished   = "{ATL-PHYS-PUB-2015-026}",
    url            = "https://cds.cern.ch/record/2037717",
    year           = "2015",
}

@Booklet{ATL-PHYS-PUB-2016-017,
    author         = "{ATLAS Collaboration}",
    title          = "{The Pythia~8 A3 tune description of ATLAS minimum bias and inelastic measurements incorporating the Donnachie--Landshoff diffractive model}",
    howpublished   = "{ATL-PHYS-PUB-2016-017}",
    url            = "https://cds.cern.ch/record/2206965",
    year           = "2016",
}

\clearpage
 
 
\begin{flushleft}
\hypersetup{urlcolor=black}
{\Large The ATLAS Collaboration}

\bigskip

\AtlasOrcid[0000-0002-6665-4934]{G.~Aad}$^\textrm{\scriptsize 99}$,
\AtlasOrcid[0000-0002-5888-2734]{B.~Abbott}$^\textrm{\scriptsize 117}$,
\AtlasOrcid[0000-0002-7248-3203]{D.C.~Abbott}$^\textrm{\scriptsize 100}$,
\AtlasOrcid[0000-0002-2788-3822]{A.~Abed~Abud}$^\textrm{\scriptsize 34}$,
\AtlasOrcid[0000-0002-1002-1652]{K.~Abeling}$^\textrm{\scriptsize 53}$,
\AtlasOrcid[0000-0002-2987-4006]{D.K.~Abhayasinghe}$^\textrm{\scriptsize 92}$,
\AtlasOrcid[0000-0002-8496-9294]{S.H.~Abidi}$^\textrm{\scriptsize 27}$,
\AtlasOrcid[0000-0002-9987-2292]{A.~Aboulhorma}$^\textrm{\scriptsize 33e}$,
\AtlasOrcid[0000-0001-5329-6640]{H.~Abramowicz}$^\textrm{\scriptsize 148}$,
\AtlasOrcid[0000-0002-1599-2896]{H.~Abreu}$^\textrm{\scriptsize 147}$
% ...
\bigskip
\\

% ...
$^{27}$Physics Department, Brookhaven National Laboratory, Upton NY; United States of America.\\
% ...
$^{33}$$^{(a)}$Facult\'e des Sciences Ain Chock, R\'eseau Universitaire de Physique des Hautes Energies - Universit\'e Hassan II, Casablanca;$^{(b)}$Facult\'{e} des Sciences, Universit\'{e} Ibn-Tofail, K\'{e}nitra;$^{(c)}$Facult\'e des Sciences Semlalia, Universit\'e Cadi Ayyad, LPHEA-Marrakech;$^{(d)}$LPMR, Facult\'e des Sciences, Universit\'e Mohamed Premier, Oujda;$^{(e)}$Facult\'e des sciences, Universit\'e Mohammed V, Rabat;$^{(f)}$Institute of Applied Physics, Mohammed VI Polytechnic University, Ben Guerir; Morocco.\\
$^{34}$CERN, Geneva; Switzerland.\\
% ...
$^{53}$II. Physikalisches Institut, Georg-August-Universit\"{a}t G\"ottingen, G\"ottingen; Germany.\\
% ...
$^{92}$Department of Physics, Royal Holloway University of London, Egham; United Kingdom.\\
% ...
$^{99}$CPPM, Aix-Marseille Universit\'e, CNRS/IN2P3, Marseille; France.\\
$^{100}$Department of Physics, University of Massachusetts, Amherst MA; United States of America.\\
% ...
$^{117}$Homer L. Dodge Department of Physics and Astronomy, University of Oklahoma, Norman OK; United States of America.\\
% ...
$^{147}$Department of Physics, Technion, Israel Institute of Technology, Haifa; Israel.\\
$^{148}$Raymond and Beverly Sackler School of Physics and Astronomy, Tel Aviv University, Tel Aviv; Israel.\\
% ...
$^{a}$ Also Affiliated with an institute covered by a cooperation agreement with CERN.\\
$^{b}$ Also at Borough of Manhattan Community College, City University of New York, New York NY; United States of America.\\
$^{c}$ Also at Bruno Kessler Foundation, Trento; Italy.\\
$^{d}$ Also at Center for High Energy Physics, Peking University; China.\\
$^{e}$ Also at Centro Studi e Ricerche Enrico Fermi; Italy.\\
$^{f}$ Also at CERN, Geneva; Switzerland.\\
% ...

\end{flushleft}


\end{document}